\documentclass[reprint,amsmath,amssymb,aps,pre]{revtex4-2}

\usepackage{graphicx}%
\usepackage{dcolumn}%
\usepackage{bm}%
\usepackage[utf8]{inputenc}
\usepackage{graphicx}%
\usepackage{hyperref}%
\usepackage{url}
\usepackage[normalem]{ulem}

\usepackage{xcolor}

\begin{document}
\title{Quality assessment of a country-wide\\bicycle node network with loop census analysis}

\author{Michael Szell}
\affiliation{Networks, Data and Society (NERDS), IT University of Copenhagen}
\affiliation{Complexity Science Hub}

\author{Anastassia Vybornova}
\affiliation{Copenhagen Center for Social Data Science (SODAS), University of Copenhagen}
\affiliation{Complexity Science Hub}

\author{Ane Rahbek Vierø}
\affiliation{Networks, Data and Society (NERDS), IT University of Copenhagen}

\begin{abstract}
    \vspace{0.3cm}
    \noindent Bicycle node networks are regional bicycle networks equipped with a wayfinding system of numbered nodes to ease recreational cycling. They spur sustainable bicycle tourism, economic spending, and local culture. Due to their country-wide scale, implementing bicycle node networks is a considerable effort and investment. Despite this investment, planning is a manual ad-hoc process that follows general design principles, but without clear performance metrics that account for the human cycling experience. Here we analyze a $28,\!215\,\mathrm{km}$ long bicycle node network spanning Denmark, developing and studying such metrics. First, a spatial analysis of geometric and topological properties  reveals high heterogeneity and local clusters of node density, face loop lengths, gradients, and feature-rich areas. Next, taking the perspective of a recreational cyclist starting at any node on the network, we create a loop census that lists all loops in the network up to day-trip length. The loop census identifies the feasible points on the network from which to take a day trip and quantifies the number of round trip choices, unveiling different levels of choice depending on the considered demographic group. While long-range cyclists can access most of the country with often overabundant choices, cyclists with stronger length and gradient limitations like families with small children can not -- which could be overcome by e-bikes. Our open-source analysis methods provide data-driven decision support for bicycle node network planning with the potential to boost the development of rural cycling and cycling tourism.\\

\end{abstract}

\maketitle

\section*{Introduction}
\begin{figure*}[t!]
   \begin{center}
   \includegraphics[width=1\linewidth]{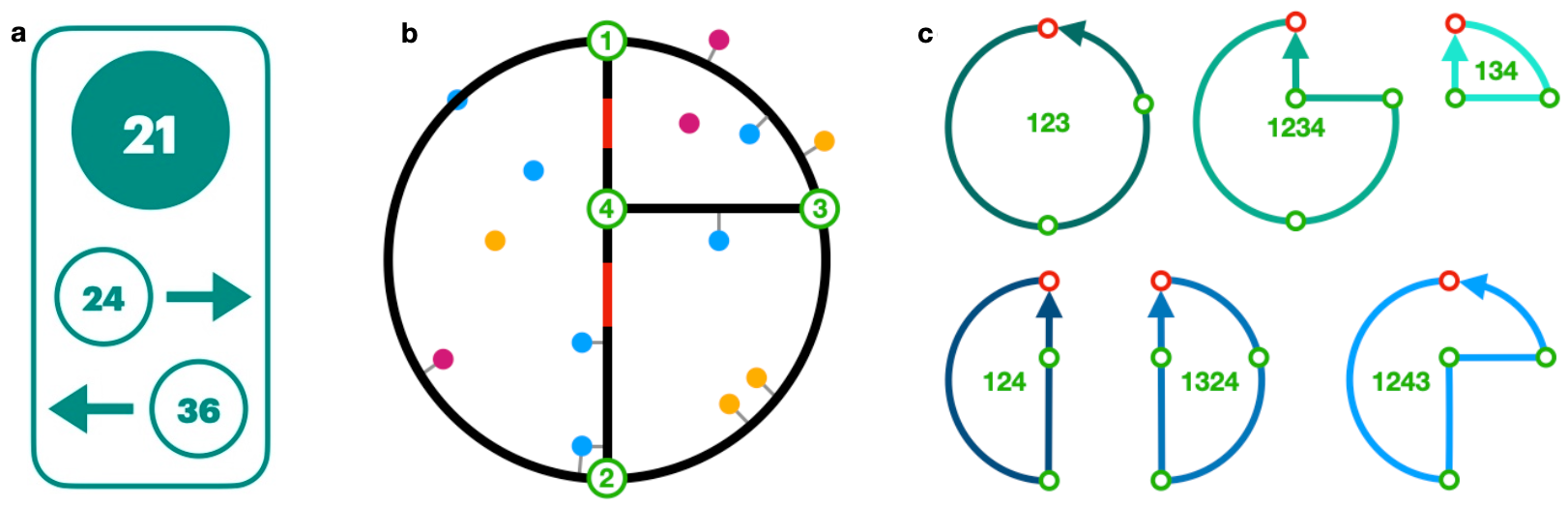}
  \caption{\textbf{Sketch of a bicycle node network (BNN), the enriched data features we have available, and the loop census method.} \textbf{a.} A typical node sign, showing the present node \textcircled{\raisebox{0.3pt}{\tiny 21}} on top, and a fork directing to the next node \textcircled{\raisebox{0.3pt}{\tiny 24}} to the right and to the next node \textcircled{\raisebox{0.3pt}{\tiny 36}} to the left. \textbf{b.} The data set comprises a bicycle network with numbered nodes spanning Denmark, with high-resolution link information such as gradient (red link segments), and points of interest (POIs) from the three categories of facility, service, and attraction (colored dots). POIs are associated to links if they are close enough (gray lines). Each POI provides water to a link (here: colored dots are close only to links $e_{1,2}$, $e_{1,3}$, $e_{2,3}$, $e_{2,4}$, and link $e_{3,4}$ but not to link $e_{1,4}$), and POIs from different categories define a link's POI diversity (here: $e_{1,2}$: 2, $e_{1,3}$: 3, $e_{1,4}$: 0, $e_{2,3}$: 1, $e_{2,4}$: 1, $e_{3,4}$: 1). \textbf{c.} For each node $i$ we construct its loop census $\mathcal{L}_i$: These are all associated loops, i.e.~all the possible round trips, without repeated links or nodes, and ignoring directionality, that a cyclist can take from node $i$. Shown is the loop census $\mathcal{L}_1$ for node 1 (red) which consists of the six loops 123, 1234, 124, 1243, 1324, 134. Loops can have different lengths, maximum gradients, and POI diversity, yielding varying suitability for different demographic groups. A planar network's face loops are given by the minimal loop basis - here the three loops 124, 134, and 234.\label{fig:overview}
  }
  \end{center}
\end{figure*}

A bicycle node network (BNN), also known as numbered-node cycle network, is a wayfinding system consisting of numbered locations (network nodes) connected via links, Fig.~\ref{fig:overview}\textbf{a,b}. It typically spans a large region or an entire country. Several countries have implemented BNNs since the 1990s \cite{dknt_studie_2021, nodemapp_cycling_2024, nederland_fietsland_knooppuntroutes_2023, province_de_luxembourg_node--node_2024} to promote recreational cycling and sustainable cycling tourism, coincidentally decreasing mass tourism's burden on the environment and climate \cite{kamb_potentials_2021, kim_does_2022}. However, planning such an extensive network comes with numerous constraining conditions and millions of potential ways to place the nodes. Despite this immense complexity, so far the development process of BNNs is entirely manual, requiring years of planning and a large amount of working power and resources. 

Research on this planning process is scarce, and the potential for computational, data-driven methods to support it is unclear. Fortunately, with the release of the Danish BNN data in 2024 \cite{caspersen_rekreative_2019, dknt_metodehandbog_2024}, we have the unique opportunity to study a country-spanning BNN quantitatively, and with local domain knowledge because we have accompanied the planning process.

Seen from a bicycle, Fig.~\ref{fig:overview}\textbf{a}, a BNN consists of signposts placed at each of the nodes directing cyclists towards the neighboring nodes on the network. This network layout enables cyclists to plan their routes with high flexibility, according to individual needs and preferences. In contrast to traditional long-haul cycling routes that only connect specific pairs of origins and destinations, it allows for a variety of round trips, adjustable trip lengths, and recreational experiences.

Taking and planning a round trip takes two main cognitive tasks: 1) Wayfinding: While taking the trip, cyclists must watch out for the next signpost, 2) Planning: Cyclists must choose one from all possible round trips. Both tasks underlie a delicate balance: First, if nodes are too close, cyclists might spend too much time on wayfinding or suffer from decision fatigue \cite{pignatiello2020decision}; but if nodes are too far apart, cyclists might become lost or unsure whether they are still on the network. Second, if there are too few possible round trips that fit the requirements of cyclists or even none, cyclists do not have enough options to choose from; but if there are too many possible options, cyclists may suffer from choice overload \cite{chernev2015choice}.

\subsection*{Implementation and previous research}
A BNN is implemented by installing signposts for cyclists' wayfinding. Therefore, the implementation does not necessarily require upgrades to the road infrastructure, which makes it potentially much cheaper to implement than e.g.~a network of protected bicycle paths. Nevertheless, for a good quality BNN it is necessary to include only road infrastructure that is suitable for recreational cycling. In addition, the BNN should offer a variation of recreational experiences; provide access to services and amenities along the way; and crucially, be safe and well-connected \cite{dknt_metodehandbog_2024}.

Despite the international success of the BNN concept, research on BNNs or on recreational cycling networks is scarce, in line with rural cycling being heavily understudied in contrast to urban cycling \cite{mcandrews_motivations_2018, kircher_cycling_2022, scappini_regional_2022, viero_network_2024}. Most literature on cycling tourism and recreational cycling focuses on organizational management, developing attractions and facilities, and only addresses the network design problem through general guidelines \cite{aschauer_guidelines_2021, caspersen_rekreative_2019, dknt_studie_2021, weston_european_2012, wirsenius_cykelleder_2021}. Meanwhile, most literature on data-driven approaches to bicycle network planning is oriented towards adding protected bicycle infrastructure in urban environments for utility cycling \cite{mauttone_bicycle_2017, caggiani_urban_2019, olmos_data_2020, szell_growing_2022, steinacker_demand-driven_2022, ospina_maximal_2022,diogo_pinto_frameworks_2026}. Therefore, up to this date, BNN planning remains a mostly manual process that requires substantial resources from planners and policy-makers. In fact, designing the Danish BNN at hand was a four 4 year long process that included 7 national stakeholders and 29 municipalities.

Only very few studies have so far aimed to develop methods specifically for recreational cycling or for BNNs. One line of approaches uses mathematical multi-criteria optimization \cite{vansteenwegen_orienteering_2011, cerna_designing_2014, malucelli_designing_2015, giovannini_cycle-tourist_2017, zhu_multi-objective_2022}. Several other studies have documented multi-criteria planning heuristics using desktop GIS \cite{derek_bicycle_2019, scappini_regional_2022}. None of the studies we are aware of account for both limitations in infrastructure and wayfinding. Also, literature on recreational cycling among children and families is scarce and focuses on environmental correlates or infrastructure features but not on network properties \cite{larouche2015built,clayton2013exploring}.

\subsection*{Our contributions}
Our research integrates geospatial analysis tools with computational methods from network science to provide a BNN assessment based on sound quantitative grounds, on both static and dynamic aspects. Using this setup, we contribute to the field threefold. 

First, we evaluate the Danish BNN, checking how its empirical properties are consistent with its stated design principles. We find mostly good overlap and regular topological properties, but also deviations, providing useful feedback for planners.

Second, we develop a generally applicable method to assess the quality of any BNN, extending the metrics from our previously developed decision support tool for BNN planning, the \texttt{BikeNodePlanner} \cite{vybornova2025bdd}. We go beyond basic static network properties and spatial features, to also consider the higher order feature of potential round trips.

Third, considering different demographic groups and their assumed requirements, we investigate three different stylized scenarios of users that would realistically want to use a BNN recreationally. Apart from a cyclist who can tolerate long trips and high gradients, stylized as ``Adult leisure cyclist'', we consider the ``Family with small children'' which has restricted length and gradient requirements, and the ``Family with e-bikes'' which can overcome range and gradient limitations \cite{rerat2021rise}. We find that adults tend to have abundant choices for round trips, but families have few choices without e-bikes.

\begin{table*}[t!]
\begin{tabular}{llll}
\textbf{Aspect}          & \textbf{Feature}                   & \textbf{Condition}                                                                                      & \textbf{Included} \\ \hline
Topological   & Link length               & \begin{tabular}[c]{@{}l@{}}Optimally 1-$5\,\mathrm{km}$\\ Maximally $10\,\mathrm{km}$\end{tabular}                       & Yes                 \\ \hline
Topological   & Face loop length          & \begin{tabular}[c]{@{}l@{}}Mostly 8-$20\,\mathrm{km}$\\ Maximally $30\,\mathrm{km}$\end{tabular}                           & Yes                  \\ \hline
Topological   & Blind end                 & Maximally 6km back and forth                                                                   & No                 \\ \hline
Topographical & Gradient                  & Maximally 6\%                                                                                  & Yes                 \\ \hline
Topographical & Flat stretches            & Maximally 500m                                                                                 & No                 \\ \hline
Service       & Distance to toilets/water & Maximally $10\,\mathrm{km}$                                                                                 & Yes                 \\ \hline
Service       & Distance to rests         & Maximally $5\,\mathrm{km}$                                                                                  & No                 \\ \hline
Service       & Distance to overnight accommodation & Maximally $40\,\mathrm{km}$                                                                                 & No                 \\ \hline
Safety        & Link traffic              & \begin{tabular}[c]{@{}l@{}}Minimize mixed traffic\\ Prioritize lower speed limits\end{tabular} & No    \\ \hline
Experience       & Route & Variation of features                                                                                 & Yes  
\end{tabular}
\caption{\textbf{The main design principles for the BNN specified by Dansk Kyst- og Naturturisme (DKNT) \cite{dknt_metodehandbog_2024}.} In our analysis we account for the most fundamental ones where data is available (Included: Yes). Most conditions permit exceptions.}
\label{tab:designprinciples}
\end{table*}

\section*{Data and methods}

\subsection*{Network data}

The Dansk Kyst- og Naturturisme (DKNT) and their project partners underwent a collaborative planning process, culminating in the publication of a Design Handbook for the BNN of Denmark in April 2024 \cite{dknt_metodehandbog_2024}, in which design principles were shaped and different communities were engaged including the authors of this paper. The raw BNN data is publicly acquirable on the GeoFA platform \cite{geofa} and regularly updated as municipalities go through their BNN developing processes. Here we use a slightly topologically simplified and preprocessed version of the raw network data fetched from GeoFA on October 11, 2024, that we make available openly \cite{zenodo}. The data preprocessing does not lose relevant content or features, see Supplementary Note~1 and Supplementary Figs.~S1 and S2 for details. The resulting network that we are analyzing has 7,170 nodes and 10,814 links with a total length of approximately $28,\!215\,\mathrm{km}$.

\subsection*{Design principles}
As a result of the planning process, DKNT and their project partners arrived at general principles for different aspects of the network and cycling experience that should hold \cite{dknt_metodehandbog_2024}. First, there is the variation of experience which specifies a variation of landscape or of cycle path features, to provide a stimulating cycling experience. Second, there should be adequately frequent accessibility and service to a variety of points of interests (POIs) such as attractions, services, or overnight accommodations. Third, on the network level, there should be safety (especially protection from vehicular traffic), connectedness, bikeability, and comfort. 

Given that our scope covers quantitative network analysis and that we do not have access to all data sets that would be necessary to tackle all mentioned aspects (such as vehicular traffic data), we do not deal with all principles, but only a subset. In particular, Table~\ref{tab:designprinciples} reports the main quantitative principles we include in our analysis (Included: Yes), from a number of quantitative and qualitative specifications, compare also Table~1 in Ref.~\cite{dknt_metodehandbog_2024}. 

\begin{figure*}[ph!]
   \includegraphics[width=0.48\linewidth]{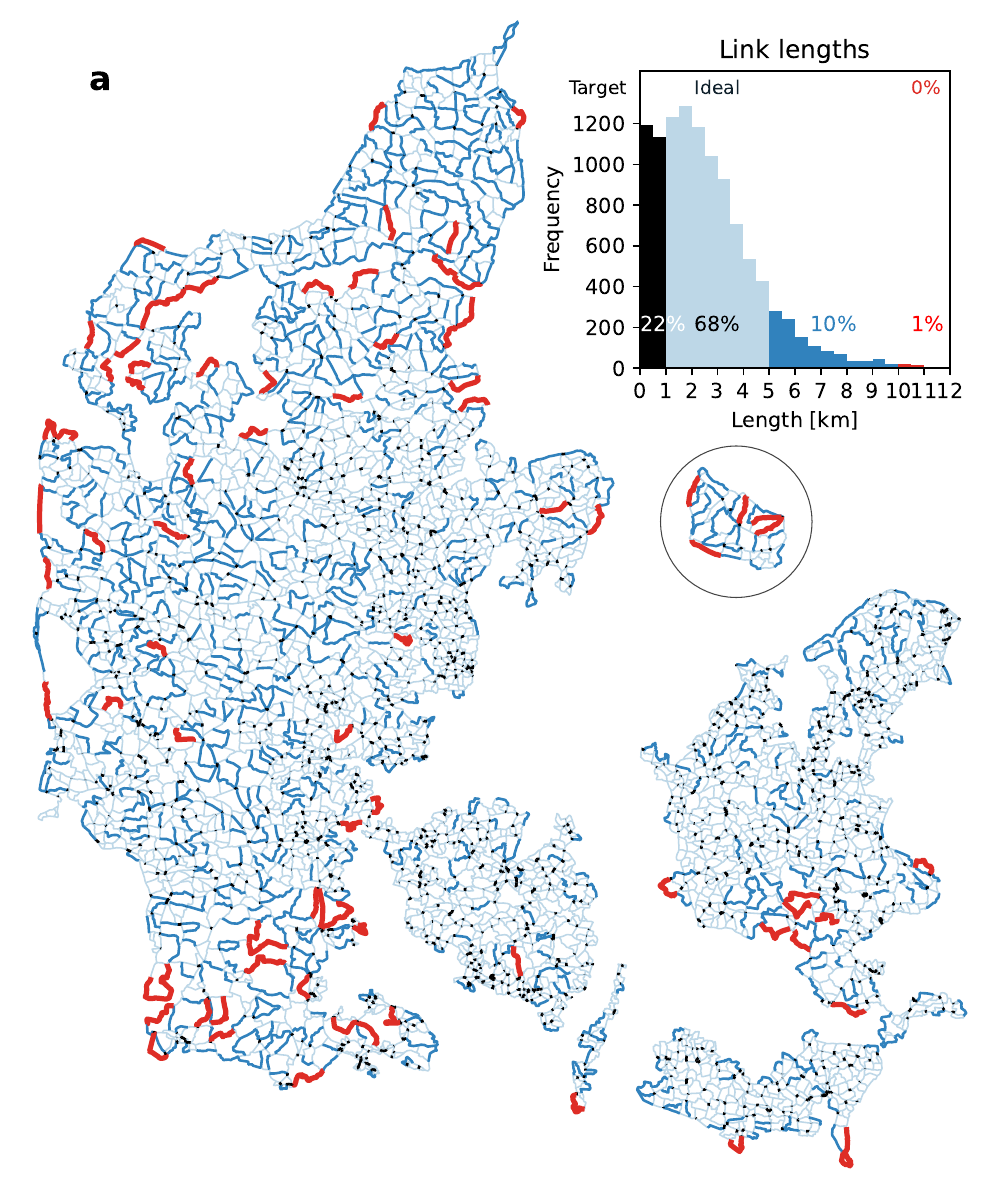}\hspace*{\fill}\includegraphics[width=0.48\linewidth]{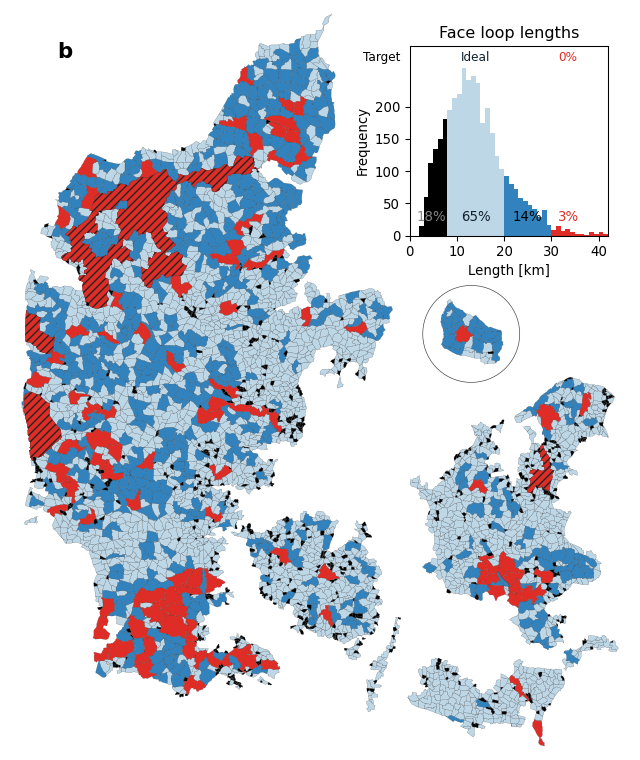}\\[18pt]
   \includegraphics[width=0.48\linewidth]{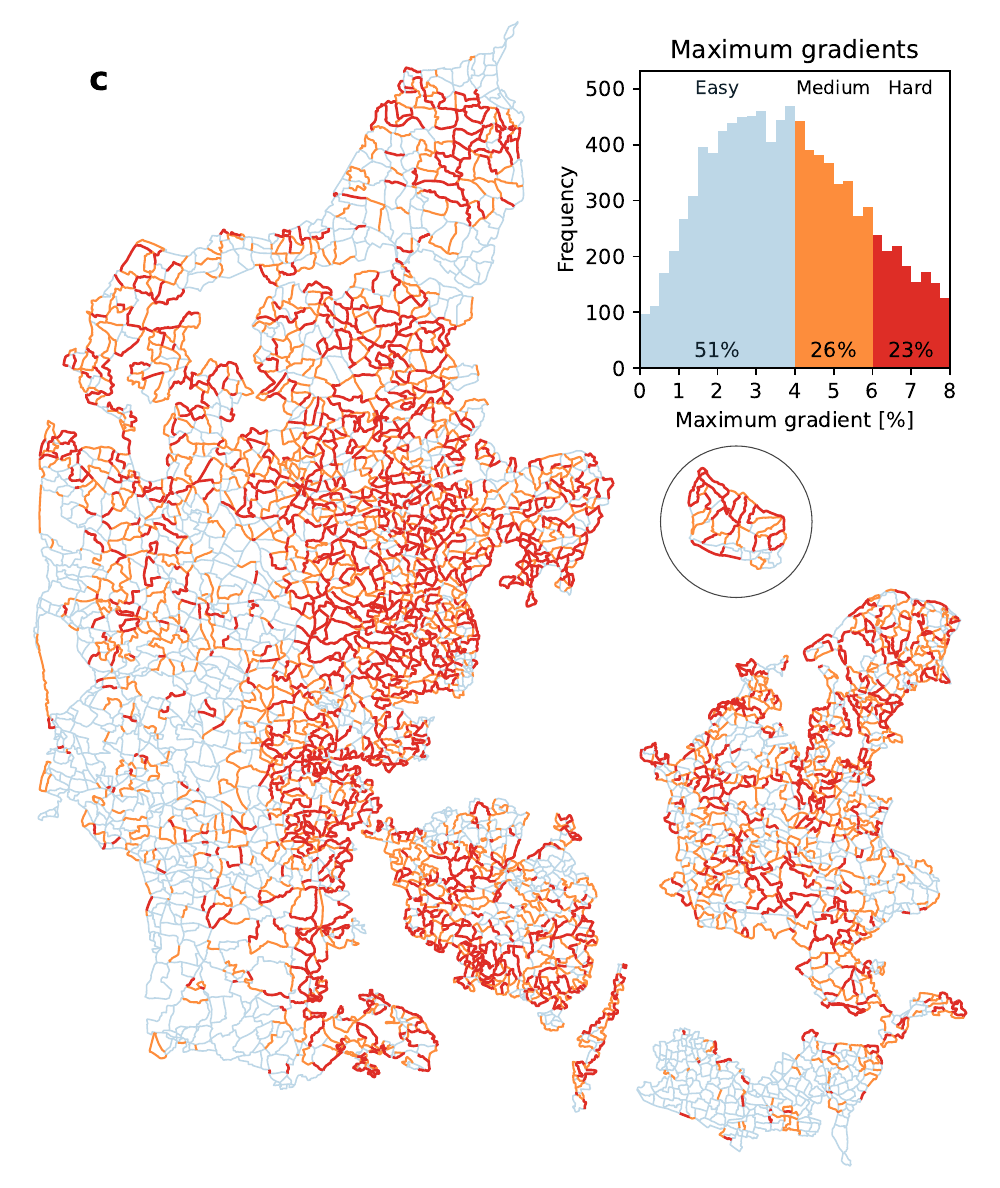}\hspace*{\fill}\includegraphics[width=0.48\linewidth]{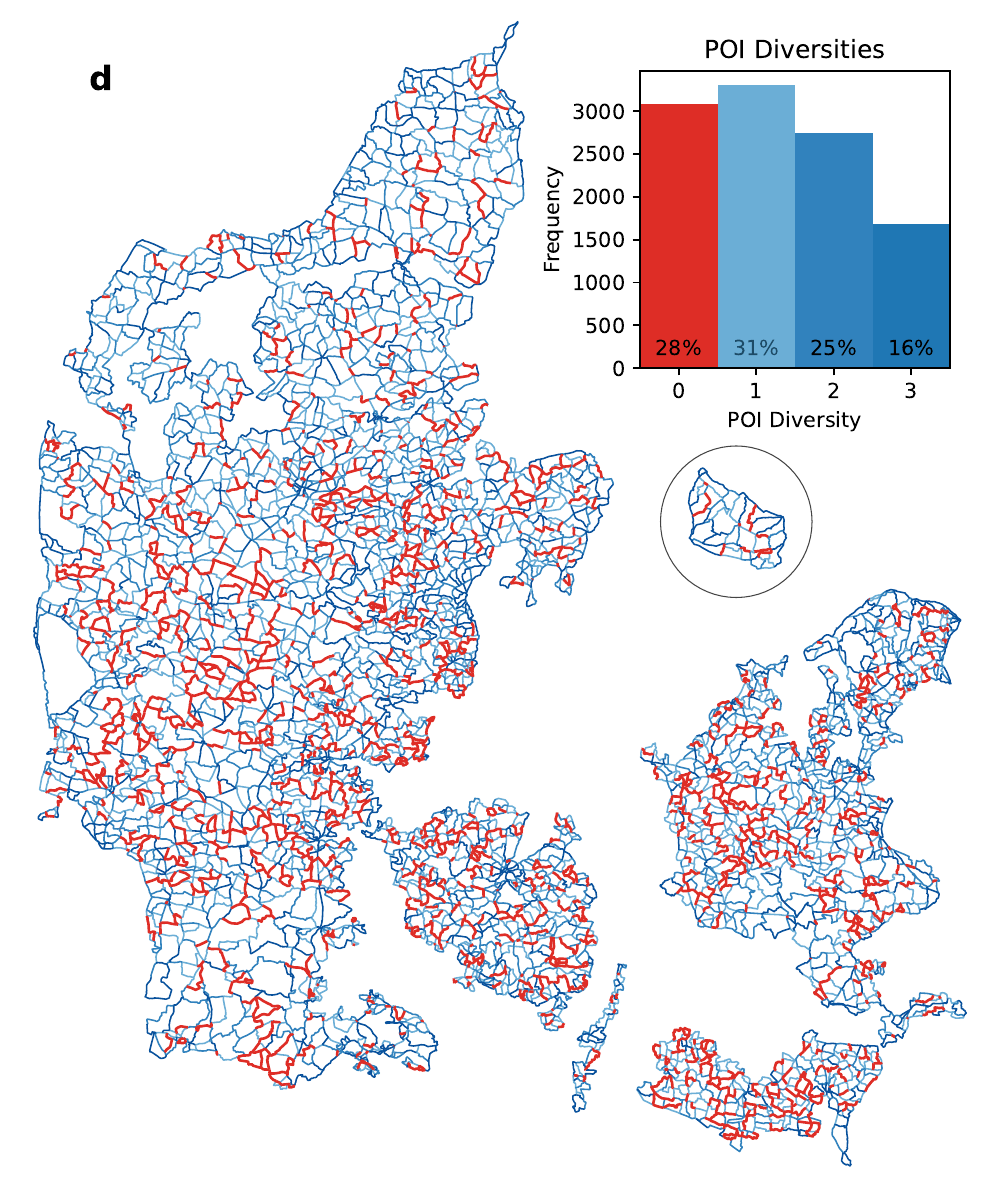}
   \begin{center}
  \caption{\textbf{Maps and distributions of geometric properties of the Danish BNN.} \textbf{a.} Link lengths, \textbf{b.} Face loop lengths (hatched faces are large bodies of water which have vacuously long shorelines), \textbf{c.} Maximum gradients, \textbf{d.} POI diversities.\label{fig:geomtriclinkproperties}
  }
  \end{center}
\end{figure*}

Out of the given design principles, we take into account the most fundamental ones: two topological (link length, face loop length), one topographical (gradient), one service related (water provision) and one experience related (POI diversity). In our analysis we first follow a basic descriptive network analysis of these features. We have previously implemented the data-driven decision support tool \texttt{BikeNodePlanner} which enables an interactive engagement with such a basic analysis for planners \cite{vybornova2025bdd}. Although we have already developed the tool and have described its features \cite{vybornova2025bdd}, we have not yet applied it for an analysis of the existing BNN. After providing such a basic analysis in this paper, we then extend this analysis with an advanced analysis based on the concept of loop census which accounts for round trips and their features. 

\subsection*{POI and elevation data}
In addition to the network data, we use POI and elevation data, fetched also on October 11, 2024 from GeoFA \cite{geofa} and from the Digital Elevation Model of Denmark \cite{dem}, respectively. We match these data to the network links to allow a filtering of results using certain criteria. We use three categories of POIs: facilities, services, and attractions, see Fig.~\ref{fig:overview}\textbf{b}. We match closeby POIs to links and define a \emph{POI diversity} $D_i \in \{0,1,2,3\}$ for each link $i$ as its number of unique POI categories. Further, we make the simplifying assumption that each POI provides a water source, which implies a \emph{water provision} $W_i$ of a link $i$ defined as $W_i = \min(1, D_i).$ In other words, any link with at least one POI is assumed to provide water to cyclists. In total, we matched 47,571 out of 54,129 POIs; the remaining POIs are too far away from the BNN. See Supplementary Note~1 for details.

\subsection*{Loop census}
The methodological novelty we develop in this paper is the loop census and its application in the bicycle network context. This method applies the graph theoretic concept of the cycle, i.e., a closed loop, to BNNs to measure the number of possible round trips at each node, which is an essential quality metric for BNNs. To not confuse the mathematical term ``cycle'' with bicycling-related terms, avoiding clunky language such as ``cycles in a cycle network'', henceforth we use the term ``loop''.

Formally, we define each node $i$'s \emph{loop census} $\mathcal{L}_i$ as the set of all loops in the network that start and end at \mbox{node $i$}. This set operationalizes all possible round trips, without repeated links or nodes, and ignoring directionality, that a cyclist can take from \mbox{node $i$}, see Fig.~\ref{fig:overview}\textbf{c}. We extend the POI diversity metric naturally from links to loops.

Using the loop census method comes with some limitations. First, all blind ends (nodes with degree 1) have an empty loop census by definition and are therefore dropped in the network preprocessing step, see Supplementary Note~1 for details. Further, due to combinatorial explosion, generally a loop census is increasing exponentially with the number of nodes considered \cite{alt1999number}. Due to such reasons of computational feasibility, here we cap all loop censuses at 30 nodes. We report in Supplementary Note~2 and Supplementary Figs.~S3 and S4 how this capping does not lose essential information.

\subsection*{Scenarios and filtering}
We run the loop census analysis for three different scenarios of stylized leisure cyclists, and for each of these scenarios for four levels of cumulative limitations, see Table~\ref{tab:scenarios}. The first scenario is ``Adult leisure cyclist'', with an assumed loop length limitation of $10\,\mathrm{km}$ to $40\,\mathrm{km}$. The second scenario is ``Family with small children'', with an assumed loop length limitation of $5\,\mathrm{km}$ to $20\,\mathrm{km}$. While $5\,\mathrm{km}$ might seem very short for a day trip, even for small children, we motivate this short lower length limit by assuming that the cyclists might incur up to $5\,\mathrm{km}$ of travel before arriving at the starting node of their round trip, and up to $5\,\mathrm{km}$ after ending it. The third scenario is ``Family with e-bikes'', with a loop length limitation of $10\,\mathrm{km}$ to $40\,\mathrm{km}$, which assumes a range doubling for e-bikes \cite{rerat2021rise}.

\begin{table*}[t!]
\begin{tabular}{c|c|c|c}
Limitation \textbackslash~Scenario                 & \textbf{Adult leisure cyclist} & \textbf{Family with small children} & \textbf{Family with e-bikes} \\ \hline
\textbf{Loop length}           & $10\,\mathrm{km}$ to $40\,\mathrm{km}$ & $5\,\mathrm{km}$ to $20\,\mathrm{km}$  & $10\,\mathrm{km}$ to $40\,\mathrm{km}$  \\
\textbf{Maximum gradient} & 6\%   & 4\% & 100\% \\
\textbf{Water source}     & every 10 km & every 10 km   & every 10 km  \\
\textbf{POI diversity}    & 3 & 3 & 3                               
\end{tabular}
\caption{\textbf{Limitations for three scenarios of diverse demographic groups.}}
\label{tab:scenarios}
\end{table*}

The first level of limitation is given by the above-mentioned length limits. For the second level, we assume the additional limitation of a maximum gradient of 6\%, 4\%, and 100\%, respectively. In other words, for adults we assume DKNT's maximal gradient of 6\%, see Table~\ref{tab:designprinciples}, while families with small children are assumed to never take a link with a maximum gradient above 4\%. However, as we assume that e-bikes make maximum gradients irrelevant, the ``Family with e-bikes'' scenario becomes equivalent to an ``Adult leisure cyclist who does not care about gradients'' scenario. The third and fourth levels of limitations are the same for all three scenarios, in line with DKNT's specifications \cite{dknt_metodehandbog_2024}: Third, a water source should be available every $10\,\mathrm{km}$; fourth, the whole round trip should have a POI diversity of 3. See Supplementary Note~3 on how we operationalize and approximate these limitations computationally.

\section*{Results}

\subsection*{Geometric and topological properties}
First, we give a descriptive overview of the network's geometric properties, including properties associated to links from enriched data, and of topological properties.

The most fundamental geometric network property for cyclists is the distribution of link lengths. Ideally link lengths should be at an intermediate distance, as too short link lengths could cause cognitive strain on having to watch out for signposts too frequently, while too long link lengths risks cyclists to become unsure whether they are still on the network, at the same time limiting the experience for demographic groups which cannot take long trips. For such reasons, the DKNT and their project partners implemented the design principles related to link length \cite{dknt_metodehandbog_2024} which should optimally be between $1\,\mathrm{km}$ and $5\,\mathrm{km}$, maximally $10\,\mathrm{km}$, see Table~\ref{tab:designprinciples}. 

The results of the link length analysis are illustrated in Fig~\ref{fig:geomtriclinkproperties}\textbf{a}. There are 10,814 links in the network, of which 68\% fulfill the ideal length requirement of being between $1\,\mathrm{km}$ and $5\,\mathrm{km}$ long. Of the remaining 32\% of links, 22\% are shorter than $1\,\mathrm{km}$, and 10\% are at the sub-optimal but acceptable length between $5\,\mathrm{km}$ and $10\,\mathrm{km}$, while 1\% of links are unacceptably long above $10\,\mathrm{km}$. Overall, the target of 0\% links above $10\,\mathrm{km}$ is almost perfectly reached, especially given that some long links cannot be avoided due to topographical circumstances. For example, several long links are found in Central and North Jutland (north west part of Denmark) along the coast or large bodies of water, where providing short links is infeasible and the risk for becoming lost is low. Nevertheless, we observe a cluster of over $10\,\mathrm{km}$ long links in Southern Denmark (south west part of Denmark) that might be avoidable. Further, the 22\% ``too short'' links, while comprising only 4\% of the total network length, could cause some unnecessary cognitive strain.

Next we analyze the face loops. These are by definition the chordless loops, therefore they provide all the round trip options in the network that cannot be shortened, see Fig~\ref{fig:geomtriclinkproperties}\textbf{b}. Their distribution and successful fit to the design principles is very similar to the link lengths. There are 3650 face loops in the network, of which 65\% fulfill the ideal length requirement of being between $8\,\mathrm{km}$ and $20\,\mathrm{km}$ long. Of the remaining 35\% of face loops, 18\% are shorter than $8\,\mathrm{km}$, and 14\% are at the sub-optimal but acceptable length between $20\,\mathrm{km}$ and $30\,\mathrm{km}$, while 3\% of face loops are unacceptably long above $30\,\mathrm{km}$. Overall, the target of 0\% face loops above $30\,\mathrm{km}$ is almost reached, especially given that some long face loops cannot be avoided due to topographical circumstances. As with the long links, a few long face loops are found in Central and North Jutland (north west part of Denmark) surrounding large water bodies, where shortcuts are not possible. In Fig~\ref{fig:geomtriclinkproperties}\textbf{b} all six such vacuously long face loops are marked with hatching. Nevertheless, again we observe a cluster of over $30\,\mathrm{km}$ long face loops in Southern Denmark (south west part of Denmark), and another cluster in the Næstved municipality in central Zealand, that might be avoidable as there are no large bodies of water there.

The last two results in this section concern link properties from enriched data. First, the distribution of maximum gradients shows that 23\% of maximum gradients are above the design condition of a 6\% gradient, see Fig~\ref{fig:geomtriclinkproperties}\textbf{c}. Despite elevations in Denmark being low compared to other countries, the country's topography still causes these gradients to appear in many parts of the country, all over northern and eastern Jutland, Bornholm, Funen, Zealand, and the Capital Region. Nevertheless, 51\% of links have a maximum gradient below 4\%, and 26\% of links are between 4\% and 6\%. The flattest regions are Lolland and western Southern Denmark. POI diversities pose a smaller and more homogeneous limitation, see Fig~\ref{fig:geomtriclinkproperties}\textbf{d}: 28\% of all links have a POI diversity of 0, 31\% have a POI diversity of 1, 25\% have a POI diversity of 2, 16\% have a POI diversity of 3. High POI diversities tend to be found in urban areas, for example around Odense or Aarhus.

\begin{figure*}[th!]
   \includegraphics{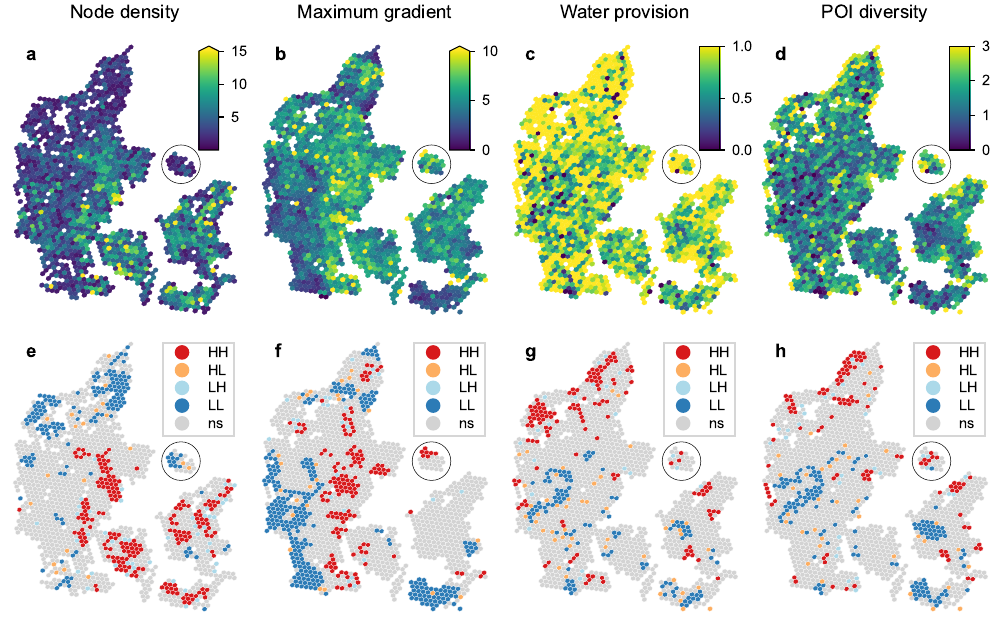}
   \begin{center}
  \caption{\textbf{Spatially aggregated properties (first row) and their local clusters (second row).} Average: \textbf{a.} Node density, \textbf{b.} Maximum gradient, \textbf{c.} Water provision, \textbf{d.} POI diversity. Maximum gradient, water provision, and POI diversity are weighted by link length. All properties display spatial heterogeneity. Local clustering (local indicators of spatial association -- LISA) of: \textbf{e.} Node density, \textbf{f.} Maximum gradient, \textbf{g.} Water provision, \textbf{h.} POI diversity. All properties display several low-low (LL) and high-high (HH) clusters.\label{fig:h3}
  }
  \end{center}
\end{figure*}

Finally, we summarize the network's topological properties, for details see Supplementary Note~4 and Supplementary Fig.~SI5. First, the network has a degree distribution sharply centered at 3. This property implies that the network's wayfinding design is consistent with simple forks which necessitate decisions only between going left or right (and no other direction) as depicted in Fig.~\ref{fig:overview}\textbf{a}. Second, the number of nodes in the network's face loops are centered around 6, implying that the network could be topologically approximated by a hexagonal grid. Assuming this model, we find that DKNT's link length specifications are consistent with DKNT's face loop length specifications \cite{dknt_metodehandbog_2024}. Such topological insights are useful for avoiding inconsistencies when setting up BNN specifications, for computationally modeling BNNs, or as a starting point for designing them algorithmically.

\begin{figure*}[t!]
   \begin{center}
   \includegraphics[width=0.48\linewidth]{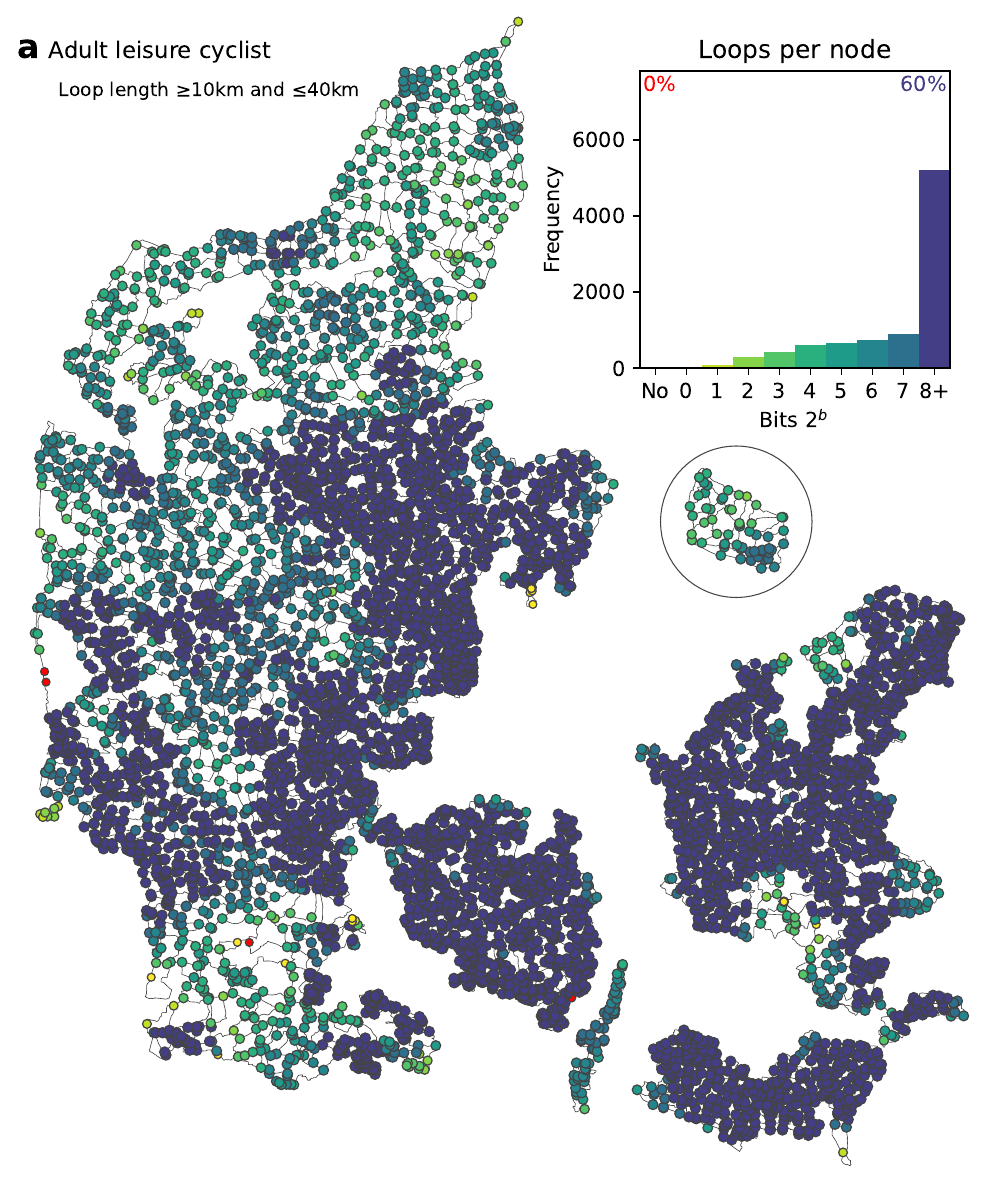}\hspace*{\fill}\includegraphics[width=0.48\linewidth]{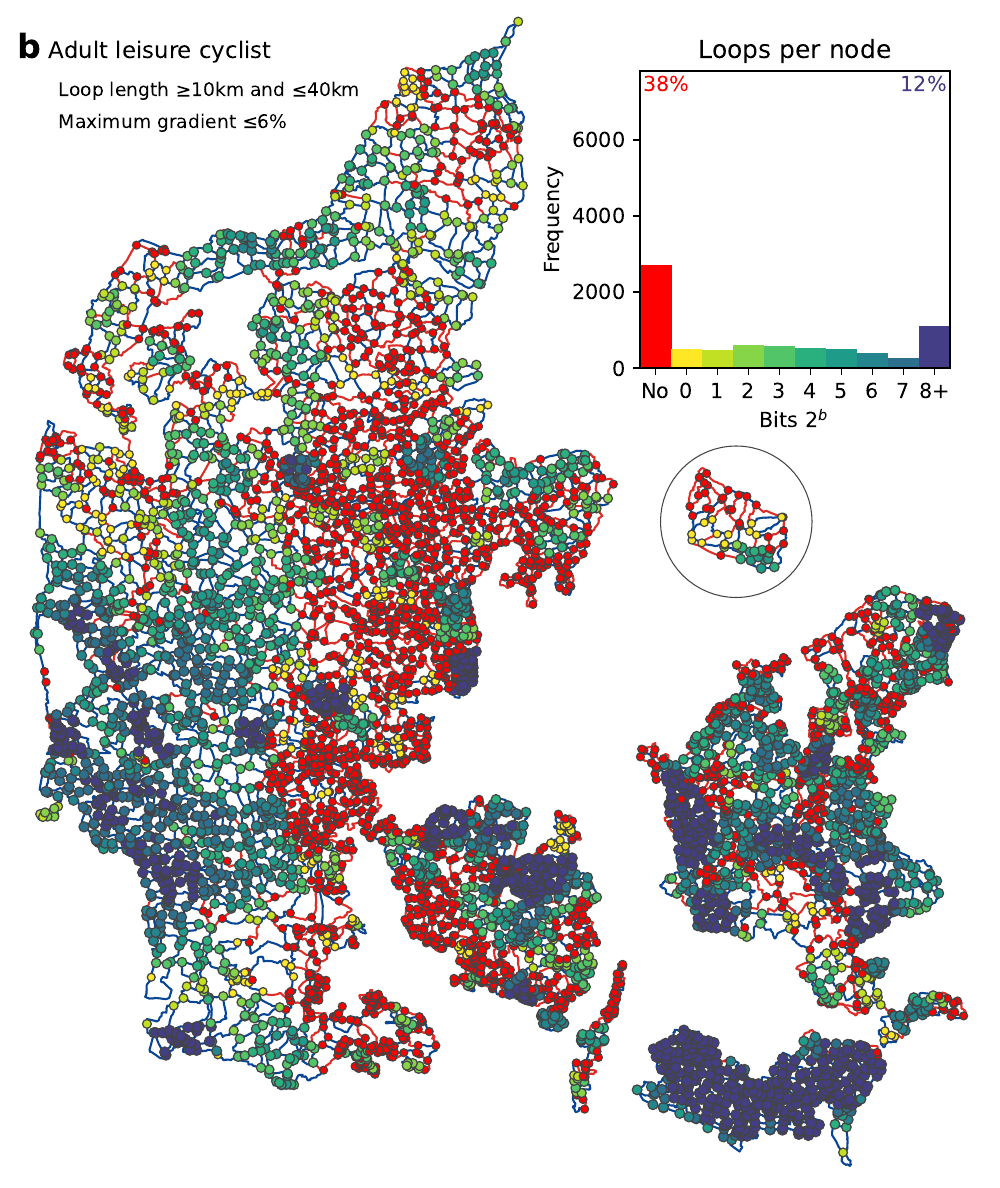}
  \caption{\textbf{Loops per node for the scenario ``Adult leisure cyclist''.} Adding restrictions reduces the loops. \textbf{a.} Restricted only by loop length $10\mathrm{-}40\,\mathrm{km}$ allows many round trips in most of the country, \textbf{b.} if also restricted by maximum gradient $\leq 6\%$ (red links show $>6\%$), then 38\% of nodes become loopless.\label{fig:scenarioadult}}
  \end{center}
\end{figure*}

\begin{figure*}[th!]
   \begin{center}
   \includegraphics[width=0.48\linewidth]{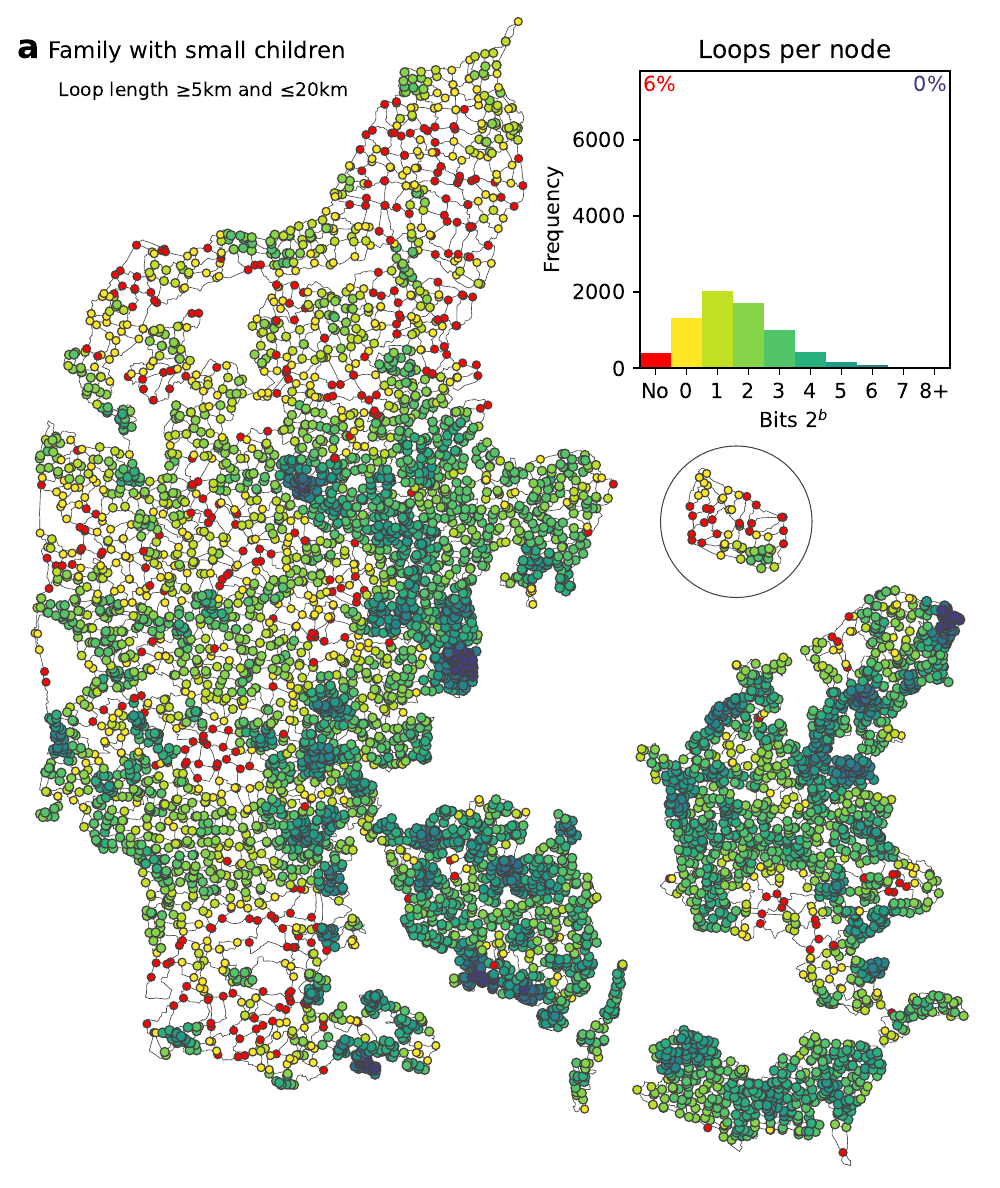}\hspace*{\fill}\includegraphics[width=0.48\linewidth]{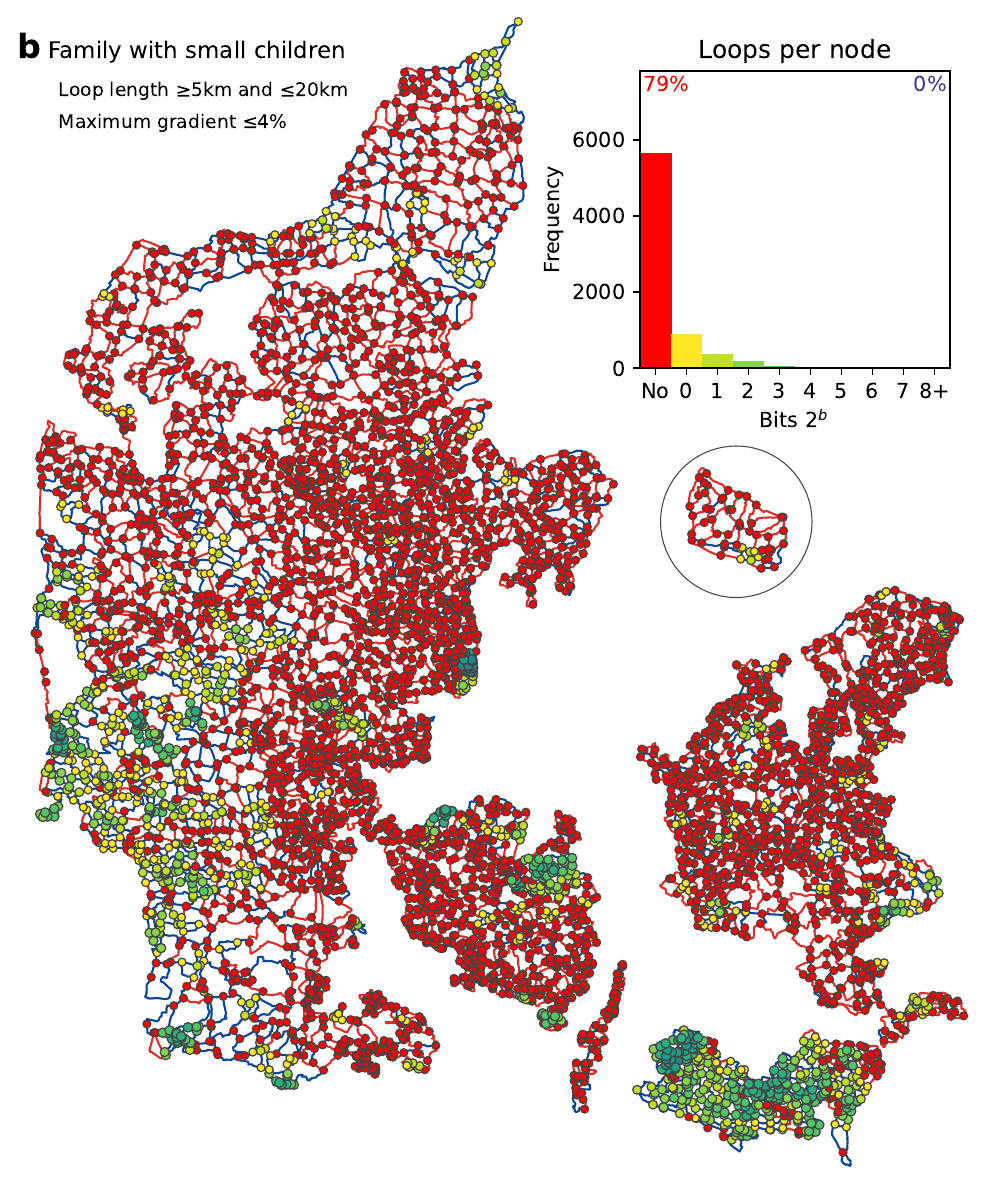}
  \end{center}
  \caption{\textbf{Loops per node for the scenario ``Family with small children''.} Adding restrictions reduces the loops. \textbf{a.} Restricted only by loop length $5\mathrm{-}20\,\mathrm{km}$ allows some round trips in most of the country, \textbf{b.} if also restricted by maximum gradient $\leq 4\%$ (red links show $>4\%$), then 79\% of nodes become loopless.\label{fig:scenariofamily}}
\end{figure*}

\begin{figure}[th!]
   \begin{center}
   \includegraphics[width=0.99\linewidth]{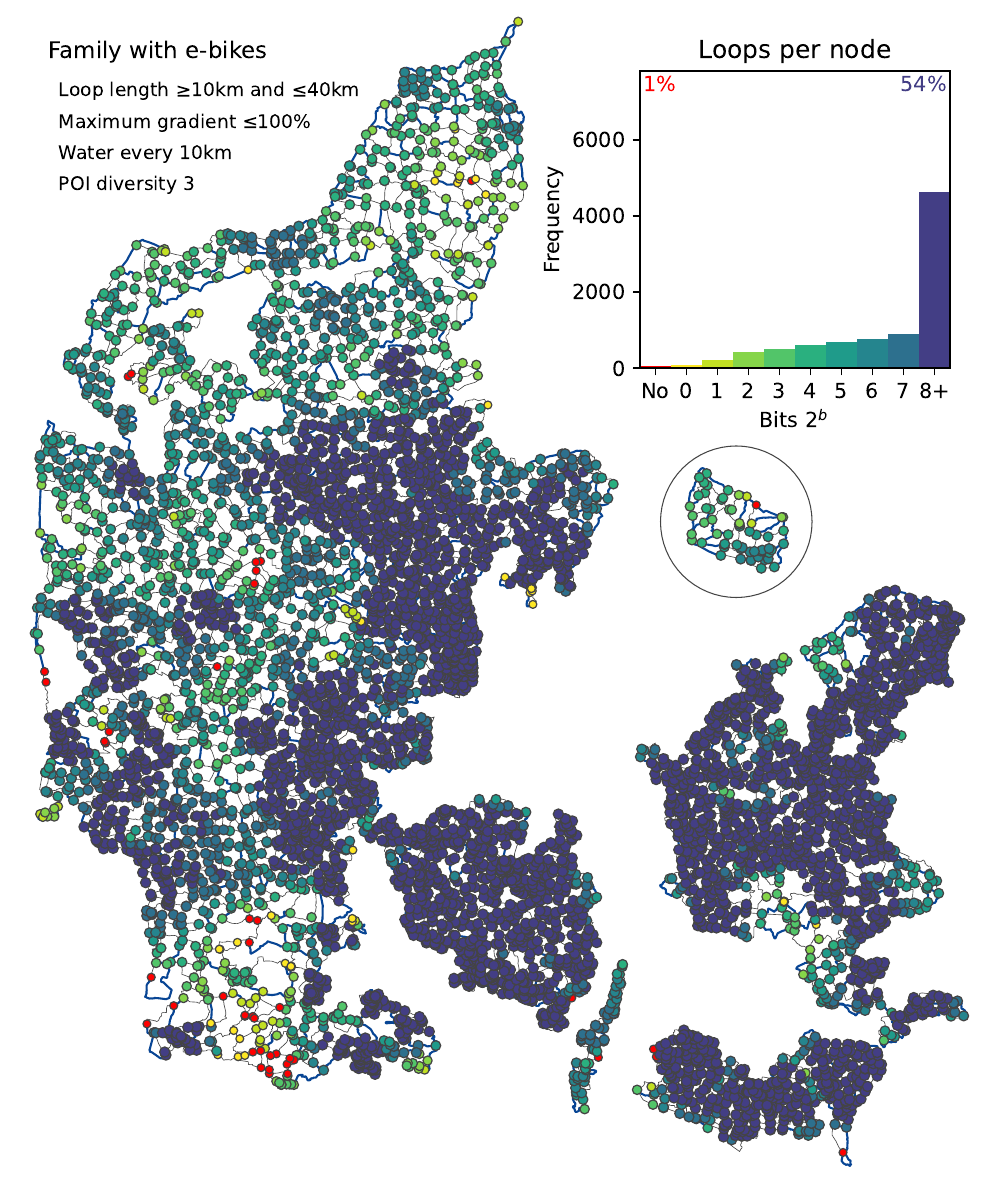}
  \caption{\textbf{Loops per node for the scenario ``Family with e-bikes''.} Adding restrictions reduces the loops, but in this scenario only negligibly. The restrictions by loop length and by maximum gradient are identical to the only length-restricted scenario of ``Adult leisure cyclist'', Fig.~\ref{fig:scenarioadult}\textbf{a}. Further restrictions: water available every $10\,\mathrm{km}$, and POI diversity 3 (blue links). Despite all these restrictions, most of the country remains well covered. \label{fig:scenariofamilyebikes}
  }
  \end{center}
\end{figure}

\subsection*{Spatial clustering}
Our basic exploration of link properties unveiled visual differences in the homogeneity of spatial distributions between maximum gradients and POI diversities, compare Fig.~\ref{fig:geomtriclinkproperties}\textbf{c} and \textbf{d}. Here we make this analysis more rigorous by applying local indicators of spatial association (LISA) on a hexagonal tesselation with contiguity neighborhoods \cite{anselin1995local}, to the spatially aggregated properties of: node density, maximum gradient, water provision, POI diversity. The idea of this spatial analysis is to provide a systematic overview to planners about both changeable properties like node density and unchangeable properties like maximum gradient, allowing to identify both potential areas and obstacles for improvement. The results are reported in Fig.~\ref{fig:h3}. The 1,524 hexagonal cells have each a side length of approximately $3.72\,\mathrm{km}$ and an area of $36.13\,\mathrm{km}^2$; this resolution was chosen to capture details adequately on both the country and regional levels.

The spatial aggregation provides a spatially regularized view onto properties, Fig.~\ref{fig:h3}\textbf{a-d}, that were previously visualized on the graph, Fig.~\ref{fig:geomtriclinkproperties}. The heterogeneity of node density is now revelaed explicitly, Fig.~\ref{fig:h3}\textbf{a}, confirming a high density on Zealand, Funen, and eastern Jutland, and a low density in the rest of Jutland and on Bornholm. Similarly, areas of high and low maximum gradients and water provision are explicitly observeable, Fig.~\ref{fig:h3}\textbf{b,c}. The aggregated view of POI diversity, Fig.~\ref{fig:h3}\textbf{d}, reveals how many coastal regions have high diversity, while some inland regions in Jutland, western Zealand and Lolland are less feature-rich. The LISA analysis, bottom row of Fig.~\ref{fig:h3}, confirms all these impressions by highlighting the specific areas which are significantly spatially clustered with high values (HH, red) and with low values (LL, dark blue).

\subsection*{Loop census analysis for different user scenarios}

We found 28,040,737 loops for up to 30 nodes and up to $40\,\mathrm{km}$ in the whole network. Henceforth we apply $\log_2$ to all loop numbers and report these as \emph{loop bits b}, with the idea that the number of loops to choose from, and the forking nature of the paths, pose a cognitive task that is naturally well described in such information theoretic terms. As an illustration, the case $b=0$ means that zero left-right decisions are possible, implying only $2^0=1$ possible round trip. Similarly, $b=1$ means that one free decision implies $2^1=2$ possible round trips, while $b=2$ means two free decisions and $2^2=4$ possible round trips, etc.

We justify the loop census analysis in Supplementary Note~5 and Supplementary Fig.~SI6 by checking that it adds relevant information on top of the node density.

\subsubsection*{Adult leisure cyclist}
We first focus on the ``Adult leisure cyclist'' scenario, Fig.~\ref{fig:scenarioadult}. Restricting only by round trip length ($10\mathrm{-}40\,\mathrm{km}$), Fig.~\ref{fig:scenarioadult}\textbf{a}, we find plenty of loops, with 60\% of nodes having 8 or more loop bits, i.e. at least 256 potential round trips to choose from. Adding the maximum gradient condition of $6\%$, Fig.~\ref{fig:scenarioadult}\textbf{b}, restricts the choices incisively, where 38\% of nodes lose all their loops, and the high 8+ bit nodes fall from 60\% to 12\%. Due to the country's topography, the biggest losses occur in hilly eastern Jutland, while the flat Lolland is mostly unaffected. Adding the water supply and then the POI diversity conditions have a smaller effect, increasing the loopless nodes to 40\%, then 42\%, and reducing the 8+ bit nodes to 10\%, then 9\%, see Supplementary~Fig.~SI7.

\subsubsection*{Family with small children}
Next we report the loop census analysis for the scenario ``Family with small children'', Fig.~\ref{fig:scenariofamily}. Restricting only by round trip length ($5\mathrm{-}20\,\mathrm{km}$), Fig.~\ref{fig:scenariofamily}\textbf{a}, we find loops on 94\% of nodes, implying that 4\% of nodes are loopless, mostly in some areas of Jutland and on Bornholm where link lengths tend to be longer. In contrast to the adult cyclist scenario, only 35 out of 7,170 nodes are 7 or 8+ bit nodes, highly concentrated in a few areas with high node density such as the summerhouse regions south of Aarhus or west of Helsingør, covering a negligible part of the country. The rest of the loop bit distribution is spread around 1-2 loop bits. Adding the maximum gradient condition of $4\%$, Fig.~\ref{fig:scenariofamily}\textbf{b}, makes 79\% of nodes become loopless, leaving most of the country without any round trip options. Adding the last two filters, Supplementary~Fig.~SI8, makes even more nodes loopless, to 80\%, and finally 88\%.

\subsubsection*{Family with e-bikes}
By construction the ``Family with e-bikes'' scenario is equivalent to an ``Adult leisure cyclist who does not care about gradients'' scenario. Therefore, for this scenario, the situation that considers only lengths and gradients is identical to the one reported in Fig.~\ref{fig:scenarioadult}\textbf{a}. Adding the last two filters reduces the loops only negligibly, moving 6\% of the 60\% of nodes from the 8+ loop bit category to lower categories, ending up with 54\% nodes having 8+ loop bits.

\section*{Discussion}

We performed a quantitative quality assessment of a large-scale bicycle node network spanning the whole country of Denmark. We accounted on one hand for basic static properties, from topological, topographical, geometric, to infrastructural. Analyzing Denmark's BNN we found general consistency with its design specifications, but also suboptimal spatial heterogeneity of properties like node density or water provision which could be future avenues for improvement. On the other hand we considered the dynamic property of the number and quality of round trips. To assess this property, we developed the loop census analysis method which is generally applicable to any network. Applying it to Denmark's BNN, assuming different restrictions for diverse demographic groups, we showed that adults tend to have abundant choices for round trips, but families with children have few choices.

Being first of its kind, our work does not fully cover additional aspects of BNNs and had to follow some simplifying assumptions. One category of simplifications followed technical reasons of computational feasibility, see Supplementary Note~4. Ideally, an algorithm for finding round trips should be relaxed to allow to some extent closed walks and circuits in addition to simple cycles, to be able to account for blind ends and other potential exemptions for repeating nodes or links. Although blind ends have only a below 3\% length share of the total network, they are often connections to important secluded POIs such as special scenic views or nature reserves along a coastline like the islands Fur or Venø.

A second category of simplifications concerns missing data or more advanced concepts. As we knew that the network provided by DKNT had already taken care of aspects of safety manually, we did not concern ourselves with it. However, with high-quality traffic data, the placement of nodes and links could be up to discussion, but would increase complexity of the analysis immensely. We can optimize all we want on the network level, but since road safety concerns dominate the uptake of cycling, especially for children and families \cite{larouche2015built,clayton2013exploring}, addressing these concerns should also be a high priority in future studies on BNNs, to support manual design already at the node placement stage. Thus, an incorporation of traffic safety into algorithmic decision support tools for BNN design like \texttt{BikeNodePlanner} \cite{vybornova2025bdd} would be highly valuable. 

Other missing aspects from Table.~\ref{tab:designprinciples} concern distances to rests, overnight accommodations, summer house areas, and other features of quality. While we took care of the variation of features via the simple POI diversity metric, also this aspect could be vastly expanded, operationalizing a ``splendor index'', taking into account land use and a more complex metric for variation of experience, such as a high diversity and \emph{frequent} variation of POIs, land use (nature, urban, historical, etc.), gradient, or shade. 

Future work could also account for the alignment of the network between neighboring municipalities \cite{dknt_metodehandbog_2024}, and for the consequences of different local approaches. In particular, some boundaries between Danish municipalities become strongly visible in some of the loops per node plots, for example Fig.~\ref{fig:scenariofamily}\textbf{a}, like Næstved which has a much smaller node density than surrounding municipalities. This discrepancy is most likely due to the design process in which municipalities can follow to some extent their own interpretation of the network design principles. It is an open future question how differences in local approaches could potentially lead to a subversion of global network goals -- a phenomenon often observed in urban bicycle networks \cite{henderson2019street,szell_growing_2022}. Nevertheless, a smaller number of possible roundtrips does not necessarily mean lower network quality -- to the contrary, it might be better to provide fewer options if those options are of higher quality, which future measures should assess better.

Importantly, in our study we also disregarded how cyclists arrive at their starting point. DKNT's focus is on summer house areas \cite{dknt_bedre_2021}, but the network should be well accessible everywhere. To avoid the indispensable arrival via unsustainable transport, i.e.~by car, the integration of the network with public transport options like rail stations should be scrutinized \cite{alessandretti2023mum}. Another aspect of integration concerns dual use of the recreational network with connecting towns and cities. While research on rural cycling is still scarce, it is clear that better rural bicycle networks are needed, especially in Denmark \cite{viero_network_2024}. This role could be partially fulfilled by recreational networks. An integration with cycle highways \cite{paulsen2024welfare} or long-distance (national or European) cycling routes \cite{pucher2008making} could generate further important synergies.

Concerning cognitive limits in wayfinding and planning, we are only aware of research on limitations for recognizing and reading signposts while riding \cite{van2020wayfinding}, but not whether decision fatigue or choice overload could play a role in the context of navigating a BNN. Research methods on navigating urban transport systems \cite{gallotti2016lost} could be adapted. It is further unclear to which extent technologies like mobile phone apps could ease or remove this strain. Therefore, we call for field experiments with and without such technologies, and for accompanying interviews, to assess the strength of such cognitive effects. If the cognitive strain is indeed an issue, our study would be relevant, for example to reduce or increase node density in the relevant regions, where possible. The preference of planning versus spontaneous trip taking should also be scrutinized for different demographic groups, where gender effects have been reported \cite{lamont2010guiding}. Further, all our scenarios and their assumptions should be validated and calibrated, for example to identify the empirical maximum gradients under different contexts.

Ultimately, network measurements like ours should not only inform, but be used for planning better networks. An open question is the fitness of use of the network to the rise of e-bike tourism \cite{van2020wayfinding,rerat2021rise}, which we started addressing through the ``Family with e-bikes'' scenario. \mbox{E-bike} network optimization involves dealing with charging stations, which is not trivial \cite{belotti2024optimization}, so our simplifications might need to be relaxed in future work. Nevertheless, we believe our study, together with the \texttt{BikeNodePlanner} tool \cite{vybornova2025bdd}, is a major first step towards a systematic approach to BNN planning. More generally, our work contributes to fostering sustainable cycling tourism and rural cycling, and our loop census method could have unforeseen applications to spatial networks where the cycle structure is important \cite{fan2021characterizing}.

\section*{Data and code}
\noindent Data: \url{https://zenodo.org/records/19222642} \cite{zenodo}
Code: \url{https://github.com/mszell/bikenwloops}

\section*{Acknowledgments}
The authors thank Kirsten Krogh Hansen and all colleagues from DKNT, DCT, Folkersma, Septima, NIRAS, Faxe Kommune, and GeoFyn, as well as Clément Sebastiao, who contributed to this work with helpful feedback, ideas, and comments. M.S. acknowledges support by the European Union through the Horizon Europe grant JUST STREETS (Grant agreement ID: 101104240). A.V. acknowledges support from Villum Fonden through the Villum Young Investigator programme (project number: 00037394).

\section*{Author contributions}
Conceptualization: MS, Data curation: MS, AV, ARV, Formal analysis: MS, Funding acquisition: MS, Investigation: MS, AV, ARV, Methodology: MS, Project administration: MS, Software: MS, Supervision: MS, Validation: MS, AV, Visualization: MS, Writing – original draft: MS, Writing – review \& editing: MS, AV, ARV

\bibliography{references_cleaned}

\end{document}


\maketitle

\noindent This is the supplementary information for the manuscript containing supplementary notes and figures.\\[1cm]

\setcounter{page}{1}
\setcounter{figure}{0}

\tableofcontents

\clearpage

\section{Data acquisition and preprocessing}\label{appendix:datasets}

\subsection*{Gradient data}
Gradient data is taken from the publicly available Digital Elevation Model (DEM) of Denmark \cite{dem} and applied to $\Delta=100\,\mathrm{m}$ segments of all links, where a segment's gradient is the slope associated with the sea level difference between its start and end point. We define a link's maximum gradient as the maximum of all gradients of its segments. This segmentation procedure ensures that a link's maximum gradient is not distorted by its length. We chose a reasonable intermediate $\Delta$ as a trade-off: Too large $\Delta$ risks maximum gradients getting lost due to averaging, too small $\Delta$ risks high maximum gradients captured but over inconsequential lengths. 

The choice of $\Delta$ is also context-dependent and should be calibrated in future work with empirical observations or feedback from cyclists. For example, for any specific $\Delta$, even if there is an unacceptably large average gradient over one or a few $\Delta$-length segments, some cyclists might be tolerant to dismounting and pushing their bicycles over such stretches.

\subsection*{POI data}
We use publicly available POI data \cite{zenodo} which was compiled from several open-source data sets (primarily OpenStreetMap and GeoFA) by the company Septima as a service to Dansk Kyst- og Nature Turisme (DKNT) within the framework of the Danish BNN project. For simplicity, here we restrict ourselves to the three top categories of facilities, services, and attractions. We use a buffer of $250\,\mathrm{m}$ to match POIs to links, where each POI can be matched to multiple links. With the matched POIs we define a \emph{POI Diversity} $D_i \in \{0,1,2,3\}$ for each link $i$ as its number of unique POI categories. Further, we make the simplifying assumption that each POI provides a water source, which we call the \emph{Water provision} $W_i$ of a link $i$ defined as $W_i = \min(1, D_i).$ In total, out of 54,129 POIs (facility: 30119, attraction: 13,881, service: 10,129) the process matched 47,571 POIs (facility: 25,283, service: 11,191, attraction: 11,097).

\subsection*{Network data}
\begin{figure*}[th!]
    \begin{center}
    \includegraphics[width=0.8\linewidth]{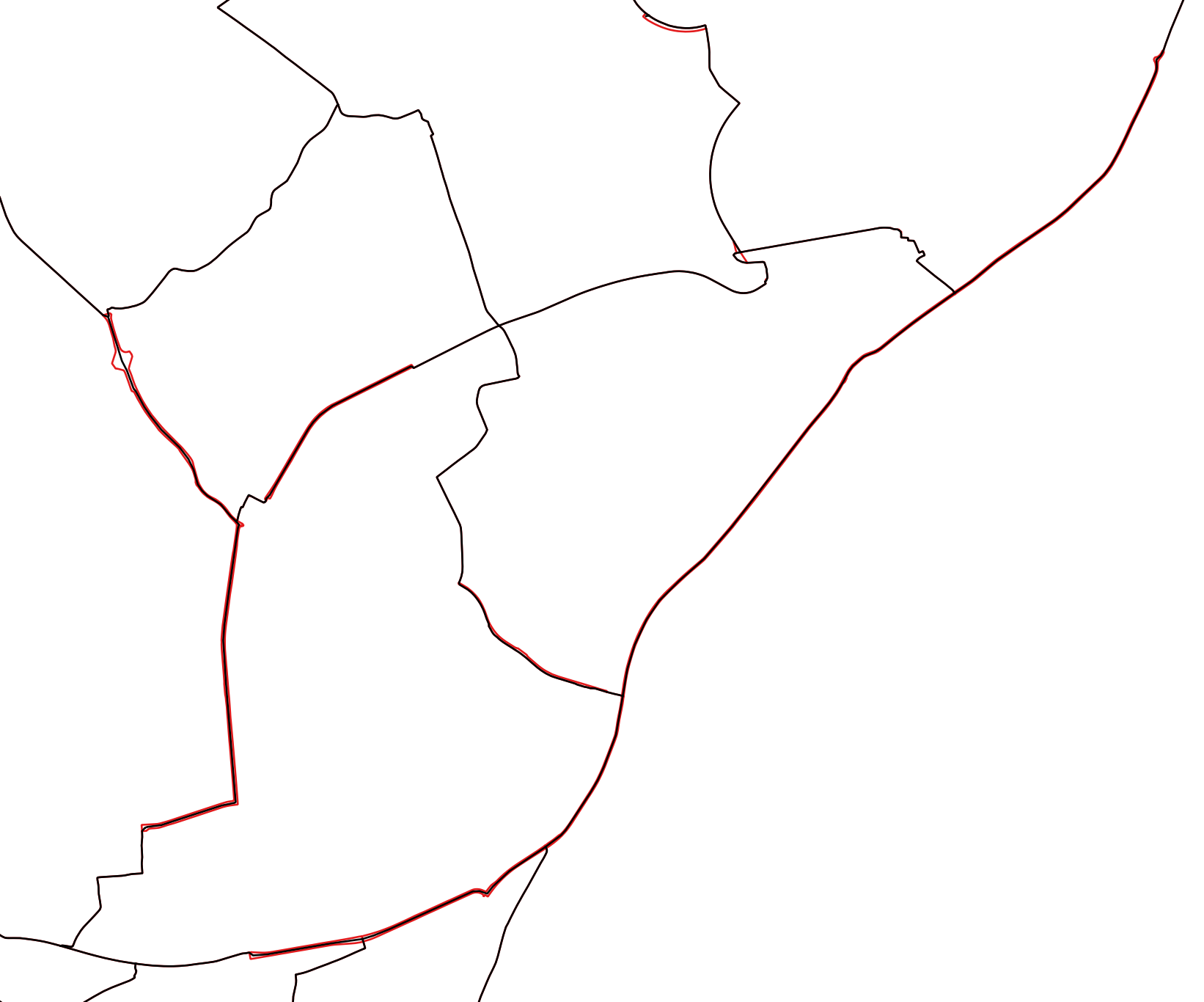}
    \caption{\textbf{Network detail demonstrating the initial transformation of dropping topological artefacts.} The resulting transformed network is in black. The raw data artefacts (parallel lines) are shown in red.\label{fig:diff_raw2neatnet}
  }
    \end{center}
\end{figure*}

Our starting point is a topologically slightly simplified version of the raw BNN data, fetched from GeoFA \cite{geofa} on October 11, 2024, constituting 25 disconnected network components of Denmark's 6 main islands Jutland, Zealand, Funen, Lolland-Falster, Bornholm, Longland, and of 19 smaller islands. The initial topological simplification is a continuity-preserving operation with a development version from October 2024 of the \verb~neatnet~ Python package \cite{fleischmann2025adaptive}, which was a necessary precondition for the subsequent analysis as it dropped topological artefacts like parallel lines -- see Fig.~\ref{fig:diff_raw2neatnet} for an example detail.

\begin{figure*}[th!]
    \begin{center}
    \includegraphics[width=1\linewidth]{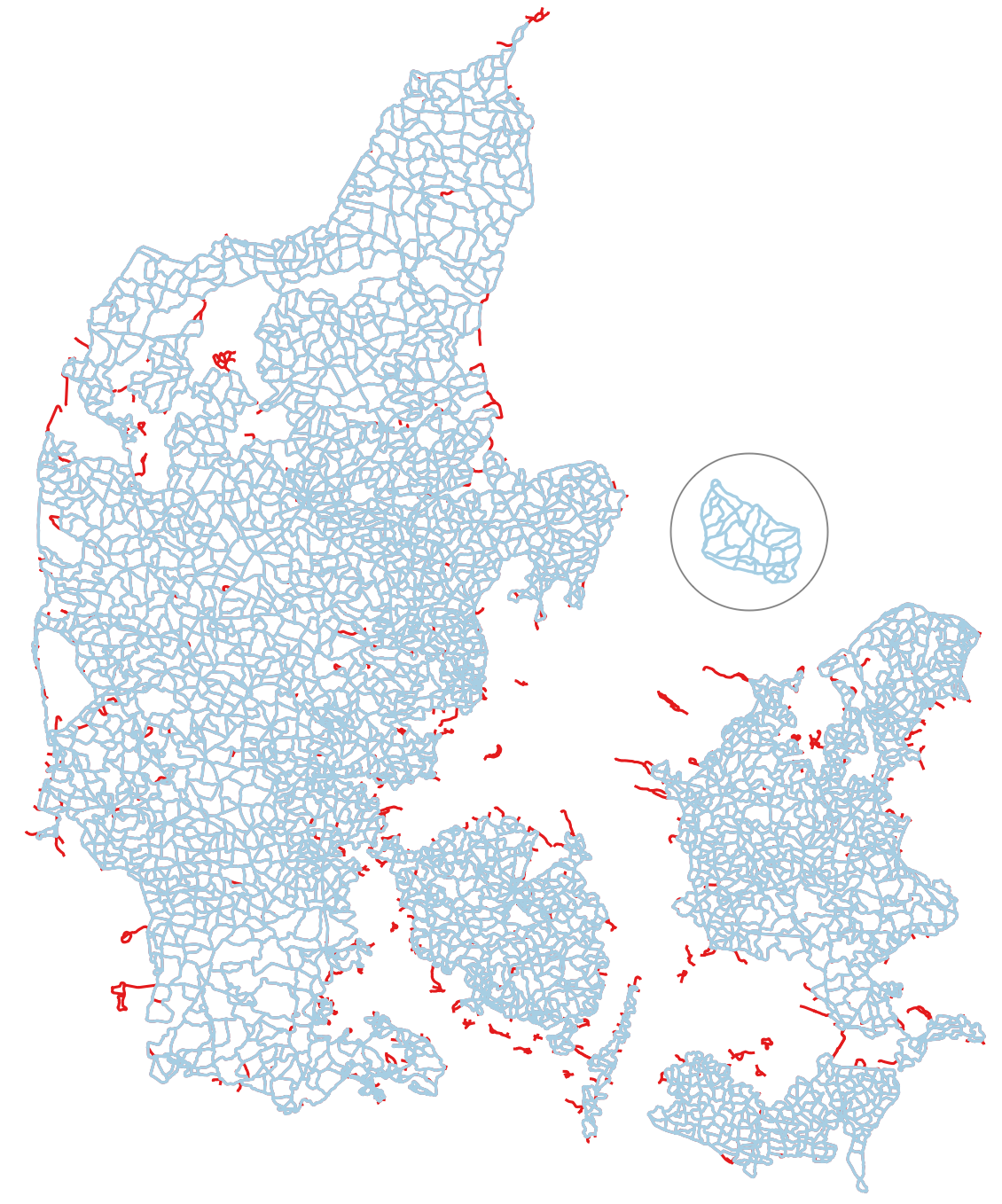}
    \caption{\textbf{Network simplification in the preprocessing step.} The dropped parts are shown in red; they consist of small disconnected components, self-loops, and blind ends.\label{fig:diff_neatnet2preproc}
  }
    \end{center}
\end{figure*}

In a preprocessing step, we then dropped the 19 smaller islands and ran for each of the remaining 6 main islands the network simplifications of dropping self-loops nad dangling nodes (blind ends). We report the dropped parts in Fig.~\ref{fig:diff_neatnet2preproc} in red. This simplification procedure was necessary to transform the network into a shape that makes a loop-centered analysis possible: First, the BNN data from the 19 dropped small network components constitute only 0.69\% of total network length and are so small that a loop-centered analysis would have been trivial as there are (almost) no loops possible. Also the analysis of link lengths and face loops would not have been meaningful due to the limited underlying topography. Second, the simplifications were necessary to drop dangling nodes and self-loops which are not relevant for our loop-centered analysis and could have caused computational issues. This preprocessing step reduced the number of connected components from 25 to 6, the number of nodes and links from 8,487 and 12,216 to 7,170 and 10,814, respectively, and the total length from $29,\!401\,\mathrm{km}$ to $28,\!215\,\mathrm{km}$. Although the number of links decreased by 11.5\%, the total length dropped was only 4.0\%.

\section{Justifying the loop capping at 30 nodes}\label{appendix:loopcutoff}
For the computational reason of exponential combinatorial explosion \cite{alt1999number}, we limit all loops to at most 30 nodes. In this supplementary note we discuss what information this capping loses, and why it is justified for our analysis.

\begin{figure*}[th!]
    \begin{center}
    \includegraphics[width=.86\linewidth]{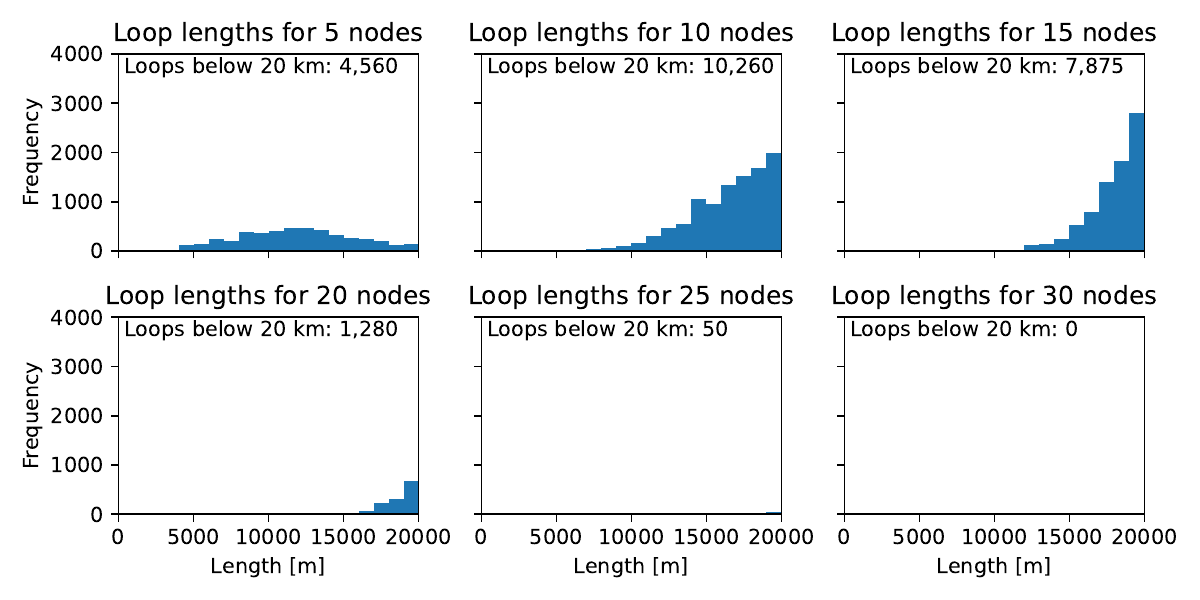}
    \caption{\textbf{Frequencies of loop lengths for different numbers of nodes.} In the range below $20\,\mathrm{km}$, the distribution shifts right with increasing number of nodes and eventually disappears above 25 nodes. Therefore, in the range below $20\,\mathrm{km}$, no loops are overlooked, and the assessment for the ``Family with small children'' scenario is exact.\label{fig:loopcapfamily}
  }
    \end{center}
\end{figure*}

First and most importantly, we check whether for the ``Family with small children'' scenario we lose any information. The concrete question is: Are there loops shorter than $20\,\mathrm{km}$ that we miss by not searching for loops with more than 30 nodes? The answer is a clear ``no'', as reported in Fig.~\ref{fig:loopcapfamily}: In the range below $20\,\mathrm{km}$, the distribution shifts right with increasing number of nodes and eventually disappears above 25 nodes. Therefore, in the relevant range below $20\,\mathrm{km}$, no loops are overlooked, and our assessment for the ``Family with small children'' scenario is exact.

\begin{figure*}[th!]
    \begin{center}
    \includegraphics[width=.86\linewidth]{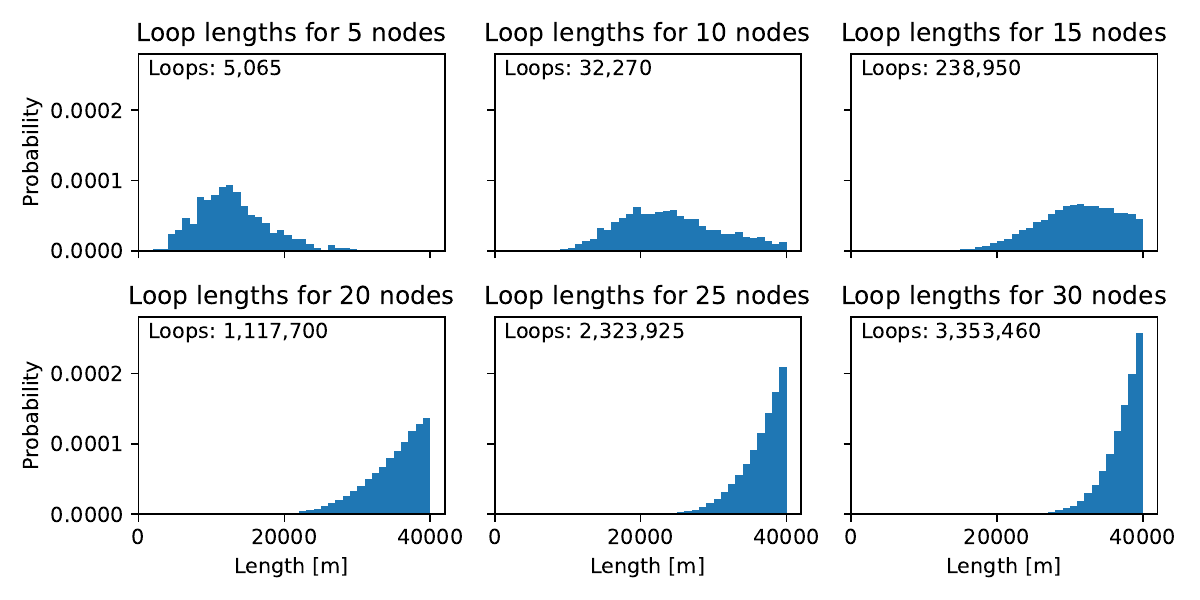}
    \caption{\textbf{Probabilities of loop lengths for different numbers of nodes.} In the range below $40\,\mathrm{km}$, the distribution shifts right with increasing number of nodes but does not disappear completely. It is likely that a few million loops are not accounted for in total, especially on the right end of the distribution close to $40\,\mathrm{km}$. Therefore, the assessments for the ``Adult leisure cyclist'' and ``Family with e-bikes'' scenarios are underestimating the true number of loops.\label{fig:loopcapadult}
  }
    \end{center}
\end{figure*}

Second, we check whether for the other scenarios, ``Adult leisure cyclist'' and ``Family with e-bikes'', we lose any information. Here we show that we indeed lose information, see Fig.~\ref{fig:loopcapadult}, as the distribution of loops shifts right with increasing number of nodes but does not completely disappear even at 30 nodes. In fact, there might be in total a few million loops with over 30 nodes but below $40\,\mathrm{km}$ not captured due to the capping. The question is whether the neglected loops are important. 

We believe this neglect is justified, for two reasons. First, the average link length is $2.61\,\mathrm{km}$, the median is $2.23\,\mathrm{km}$. A loop with over 30 nodes but shorter than $40\,\mathrm{km}$ would need to consist of 30 links below the 0.29-quantile of $1.34\,\mathrm{km}$, on average, which is improbable. Also from a practical planning perspective, it is unlikely that cyclists would want to select such a large number of nodes for their round trips. Second, for the ``Adult leisure cyclist'' scenario, the number of round trips per node is dominated by the number of links in its neighborhood that have a gradient above 6\%. Therefore, most neglected loops have to be in areas with both few such high gradient links and high node densities. However, these are at the same time areas where there is already a high saturation of loops at 8+ loop bits, for example Lolland-Falster. As 8+ is our maximum loop bit class, adding more loops to an 8+ loop bit node would not change anything for our analysis. We therefore believe that the neglect of such many-node loops would not change the results substantially, at least not qualitatively. Of course, this claim should be checked in future work, to whatever extent made possible by computational or algorithmic advancements.

\section{Computational limitations and approximation of water provision}\label{appendix:operationalization}

 Our implementation is limited by the use of networkX's algorithm \verb|simple_cycles()| for generating all simple cycles, which is relatively slow and cannot natively take into account spatial properties from our enriched data such as POIs as it is not developed for data-enriched, spatial networks. Further, it does not allow for exceptions for closed walks and circuits in addition to simple cycles, to be able to account for blind ends and other potential exemptions for repeating nodes or links. Its only parameter is \verb|length_bound|, which allows for a node-length cutoff, but any additional considerations had to be manually coded, for simplicity in Python, which is slow. Although python-igraph implemented its \verb|simple_cycles()| routine in summer 2025, which should be faster than the networkX algorithm, the same limitations persist there. A custom extension of the core algorithms in a low level language like C or Rust would be necessary for the needed speed improvement, which would have been a too involved software engineering task outside the scope of the paper.

 Because of above limitations, we were not able to calculate the specific loop property of water provision exactly but needed to resort to an approximation here: Our implementation assumes that whenever a link has a water source, then this water source is located on the center of the link. The inaccuracy introduced by this approximation cancels out on average, but it could introduce some false negatives (water source calculated after slightly above $10\,\mathrm{km}$ when in reality below) with the same magnitude of false positives (water source calculated after slightly below $10\,\mathrm{km}$ when in reality above). In the worst case, a point of water provision could be shifted half of the length of a link. For median link lengths of $2.23\,\mathrm{km}$ the introduced worst possible inaccuracy is $2.23\,\mathrm{km}$ (half the link length on both ends of a waterless segment). Nevertheless, assumptions about water provisions are already approximate: every POI is assumed to provide a water source, POIs are matched at a distance of $250\,\mathrm{m}$, and the $10\,\mathrm{km}$ maximum distance is an arbitrary normative specification. Therefore, we believe that adding a bit more leeway to the calculation of the $10\,\mathrm{km}$ water provision requirement does not change results substantially. A possibility to account for these assumptions conservatively could be to calculate with a reduced maximum distance, changing the $10\,\mathrm{km}$ to a lower value.

\section{Topological properties}\label{appendix:topologicalproperties}
The bicycle node network (BNN) of DKNT is a manually designed network. In a wayfinding context, BNNs work best when their nodes are simple forks that necessitate decisions only between going left or right (and no other direction). In short, nodes should have degree 3. Further, the choice of link length limits the possible lengths of face loops and vice versa. Since the stated design principles specify optimal lengths for both, here we check if these specifications are consistent with each other. Next, we measure the empirical topological properties of the designed network, to see how it fits the specifications. Last, here we also aim to understand which regular network could be the best model for a BNN: If there were no geographic constraints, what would be the ``platonic ideal'' of a BNN? Can we infer implicit design principles from studying the empirical network?

\begin{figure*}[th!]
   \begin{center}
\includegraphics{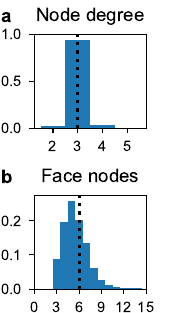}\qquad
   \includegraphics[width=0.24\linewidth]{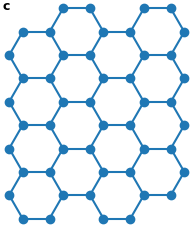}\qquad \quad\includegraphics[width=0.21\linewidth] {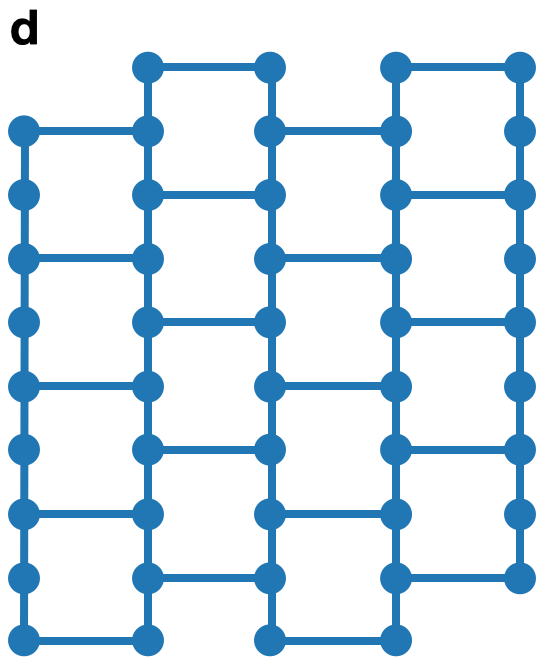}
  \caption{\textbf{Topological properties and models for the BNN.}  \textbf{a.} Degree distribution: The degree of most nodes is 3 (dotted line), implying the prevalence of simple forks where cyclists can choose only between going left or right (and no other direction). \textbf{b.} Distribution of nodes per face: The empirical network does not have a constant number of nodes per face, but the distribution is centered around 6 (dotted line). \textbf{c.} Given the degree, link length, and face loop length specifications, the network could be modeled as a hexagonal graph. \textbf{d.} Alternatively, the network could be modeled as a staggered square grid, isomorphic to the hexagonal graph.\label{fig:syntheticnets}
  }
  \end{center}
\end{figure*}

We find that the empirical network's degree frequencies are: 0 nodes of degree 0 or 1, 139 nodes of degree 2, 6775 nodes of degree 3, 255 nodes of degree 4, 1 node of degree 5. These frequencies yield a degree distribution of $P(2)=1.9\%, P(3)=94.5\%, P(4)=3.6\%, P(5)=0.01\%$. Therefore, the vast majority of nodes has indeed degree 3, which is consistent with our intuition that nodes should be as simple as possible forks for cyclists, see Fig~\ref{fig:syntheticnets}\textbf{a}. Studying the distribution of nodes per face loop, we also find it centered around 6, albeit with higher variation than the degree distribution, Fig.~\ref{fig:syntheticnets}\textbf{b}.

Given that each node should have degree 3, and that many face loops have around 6 nodes, the most natural model for a BNN is a hexagonal network, see Fig.~\ref{fig:syntheticnets}\textbf{c}. Plugging in the targeted optimal link lengths of 1-$5\,\mathrm{km}$ yields optimal face loop lengths of 6-$30\,\mathrm{km}$, which is a slightly larger range than the specified target of 8-$20\,\mathrm{km}$ \cite{dknt_metodehandbog_2024}, but consistent. It is possible to achieve both specifications exactly though, for example with a staggered square grid, see Fig.~\ref{fig:syntheticnets}\textbf{d}. It is isomorphic to the hexagonal network, however, link lengths are not uniform here but distributed such that $\frac{1}{3}$ of all links (the horizontal links in Fig.~\ref{fig:syntheticnets}\textbf{d}) are twice as long as the rest (the vertical links in Fig.~\ref{fig:syntheticnets}\textbf{d}). Varying link lengths from 1-$5\,\mathrm{km}$ in the staggered grid leads to face loop lengths of 8-$20\,\mathrm{km}$, as specified.

\section{How well does node density explain loop bits?}\label{appendix:densityproxy}
Given the circumstantial visual evidence from the loop census analysis that more loops sometimes coincide with more nodes per area, we want to check more rigorously whether or to which extent node density is a good proxy for loop bits. Intuitively it makes sense that more nodes per area can imply more links per area, which would imply more loops. Also, if node density were a good proxy for loop bits, then the costly loop census analysis would not be necessary to perform, and node density could serve as a ``good enough'', cheap estimate for determining round trip options.

\begin{figure*}[th!]
   \includegraphics{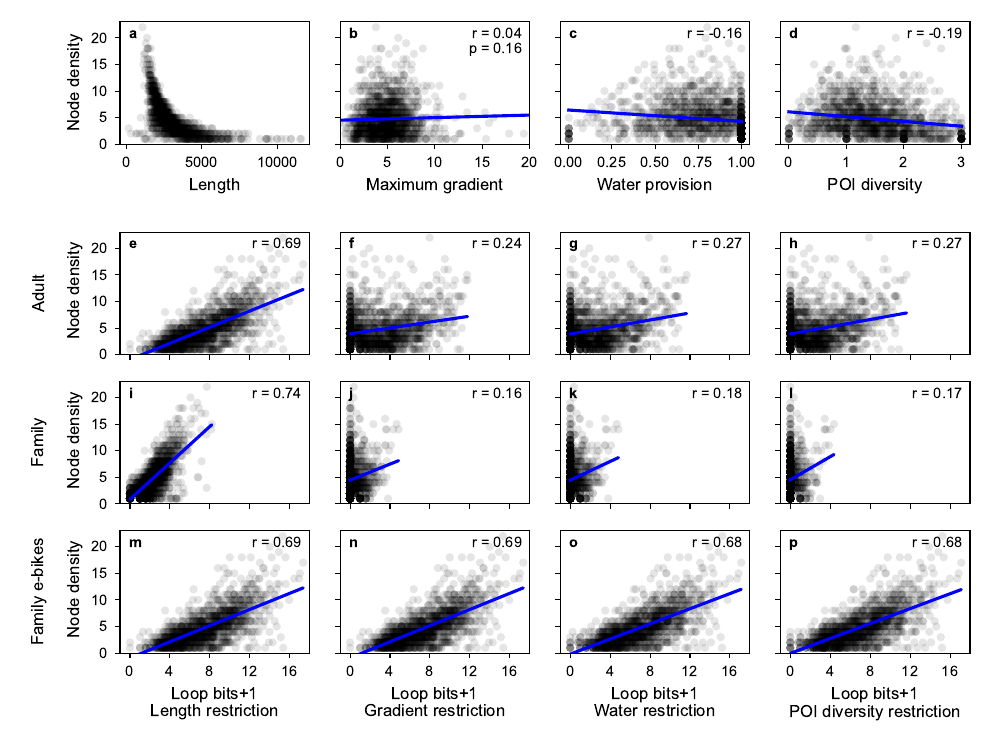}
   \begin{center}
  \caption{\textbf{Node density versus properties.}\label{fig:nodedensityscatter}
  }
  \end{center}
\end{figure*}

Prior to checking this proxy hypothesis, we assess potential correlations with topographical attributes, performing linear regressions of node density versus geometric link properties (maximum gradient, water provision, POI diversity). To be able to perform regressions, we spatially aggregate these properties via averages on a hexagonal H3 grid with resolution 6, yielding 1,524 grid elements with approximately $3.72\,\mathrm{km}$ side length and $36.13\,\mathrm{km}^2$ area. Figure \ref{fig:nodedensityscatter}\textbf{a}--\textbf{d} reports the regressions between node density and average property values over all grid elements. The only relationship we do not correlate is node density versus link length, as this is a clearly non-linear negative association, Fig.~\ref{fig:nodedensityscatter}\textbf{a}. Given the topological properties of the BNN and its planarity, this association is intuitive, as more nodes imply that there is less space for links. For the regression on the maximum gradient we find the smallest and only non-significant association, at $r=0.04$ and $p=0.16$, implying that node density and maximum gradient are not associated linearly. Water provision and POI diversity are significantly associated negatively, with $r=-0.16$ and $r=-0.19$, respectively, and $p<10^{-9}$.

Finally, to check the hypothesis of node density being a proxy for loop bits, we perform the same regressions as before but for node density versus loop bits over all the three scenarios and their cumulative restrictions. We find all these associations to be positive and significant (all $p<10^{-9}$), but with varying strengths. As one could expect, the least restricted cases are most strongly associated, Fig.~\ref{fig:nodedensityscatter}\textbf{e},\textbf{i},\textbf{m}, at $r=0.69$ and $r=0.74$, respectively. This association could be interpreted as evidence that node density is a good proxy for loop bits, thus for round trips. However, once the maximum gradient restriction is added in the first two scenarios, Figs~\ref{fig:nodedensityscatter}\textbf{f},\textbf{j}, the association breaks down to $r=0.24$ and $r=0.16$, respectively. The distribution of gradients thus destroys many correlations, weakening the power of node density as a proxy. Further restrictions naturally push the loop bits slightly towards zero, Figs.~\ref{fig:nodedensityscatter}\textbf{g},\textbf{h},\textbf{k},\textbf{l}, and the correlations increase negligibly, to $r=0.27$, $r=0.27$, and $r=0.18$, $r=0.17$, respectively. Examining the third scenario, Fig.~\ref{fig:nodedensityscatter}\textbf{m}--\textbf{p}, reveals that most of the correlation collapse comes indeed from the spatial distribution of the maximum gradients, because the water and POI diversity restrictions do not substantially lower the correlations, from $r=0.69$ only down to $r=0.68$. To summarize, without a maximum gradient restriction, node density can be a reasonably good proxy for loop bits, but with a maximum gradient restriction the correlation mostly breaks down, at least for Denmark.

\clearpage
\renewcommand{\thesection}{Supplementary Figures} 
\section{}


\begin{figure*}[th!]
   \begin{center}
   \includegraphics[width=0.48\linewidth]{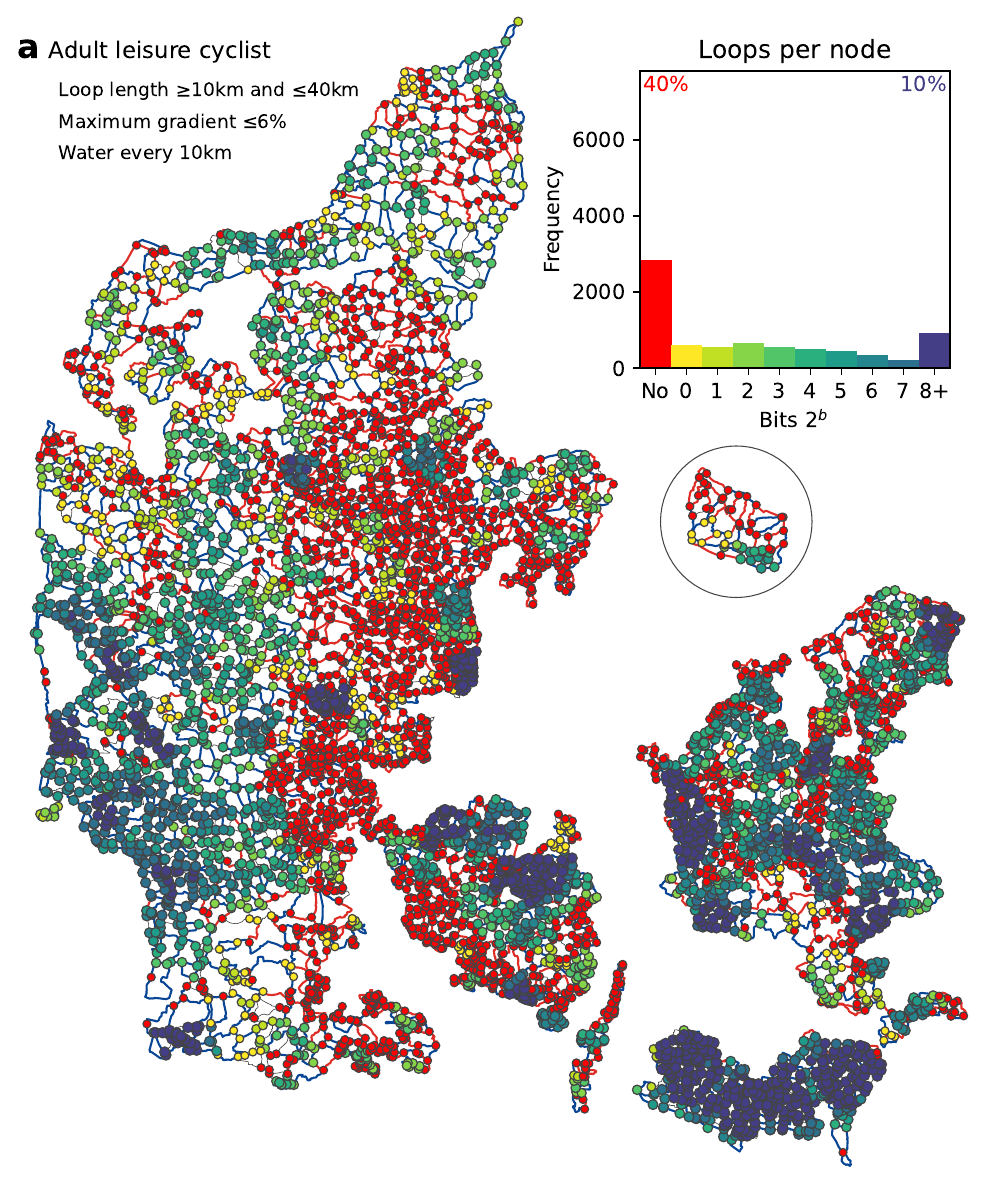}\hspace*{\fill}\includegraphics[width=0.48\linewidth]{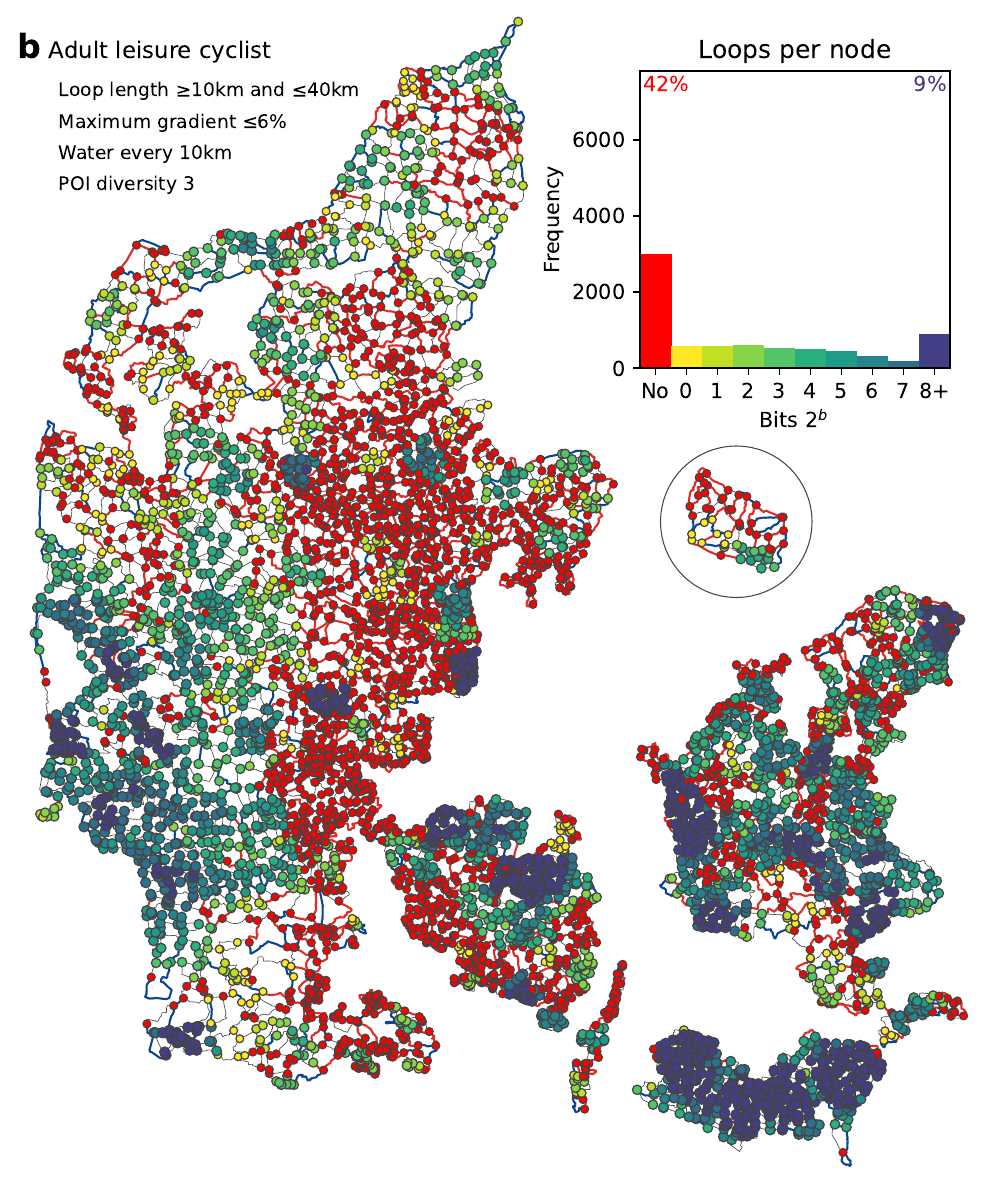}
  \caption{\textbf{Loops per node for the scenario ``Adult leisure cyclist''.} Adding restrictions on top of length and maximum gradient reduces the loops further. \textbf{a.} Also restricted by water available every $10\,\mathrm{km}$ (blue links), \textbf{b.} also restricted by POI diversity 3 (blue links).\label{fig:scenarioadultwaterpoi}}
  \end{center}
\end{figure*}

\begin{figure*}[th!]
   \begin{center}
   \includegraphics[width=0.48\linewidth]{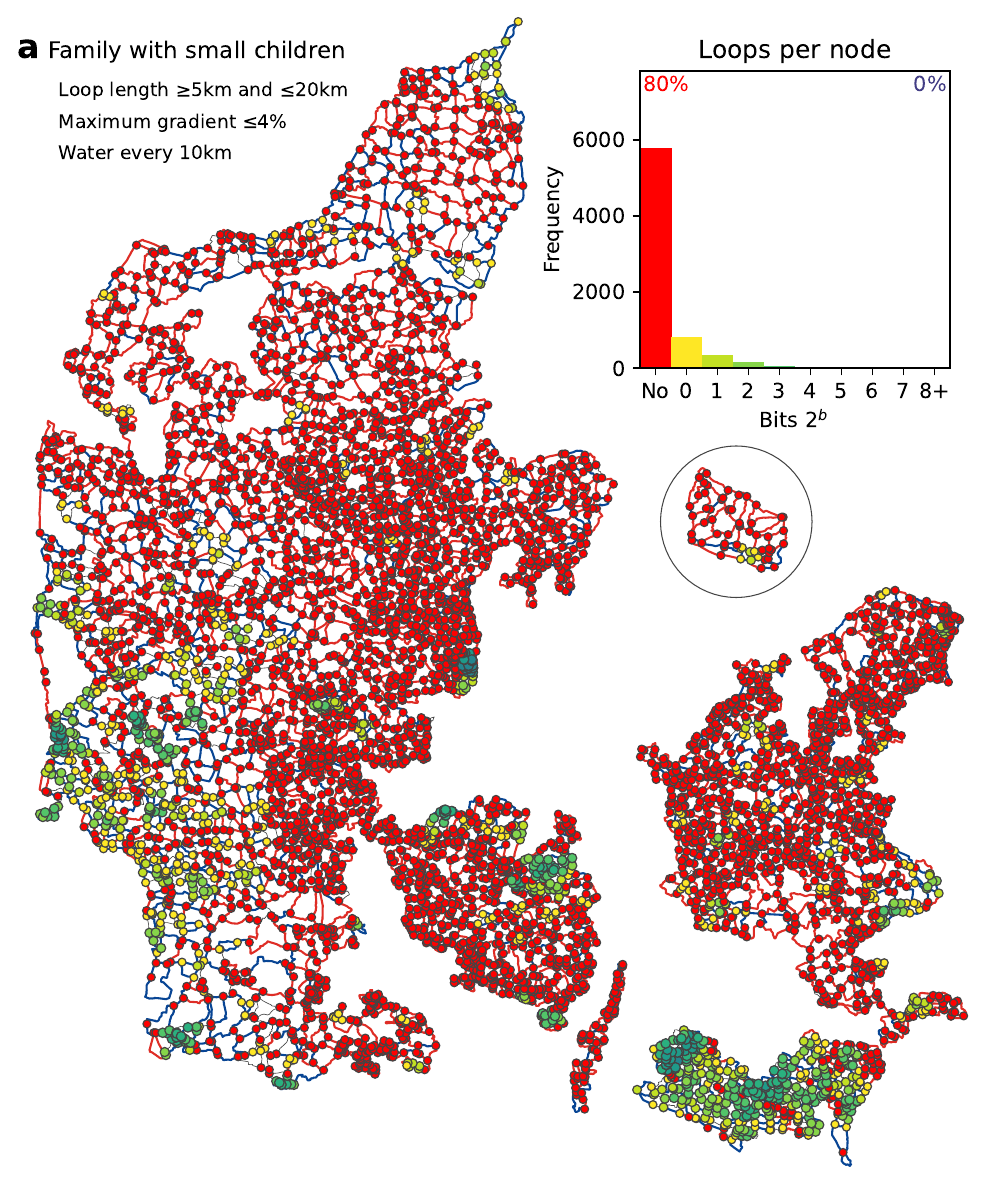}\hspace*{\fill}\includegraphics[width=0.48\linewidth]{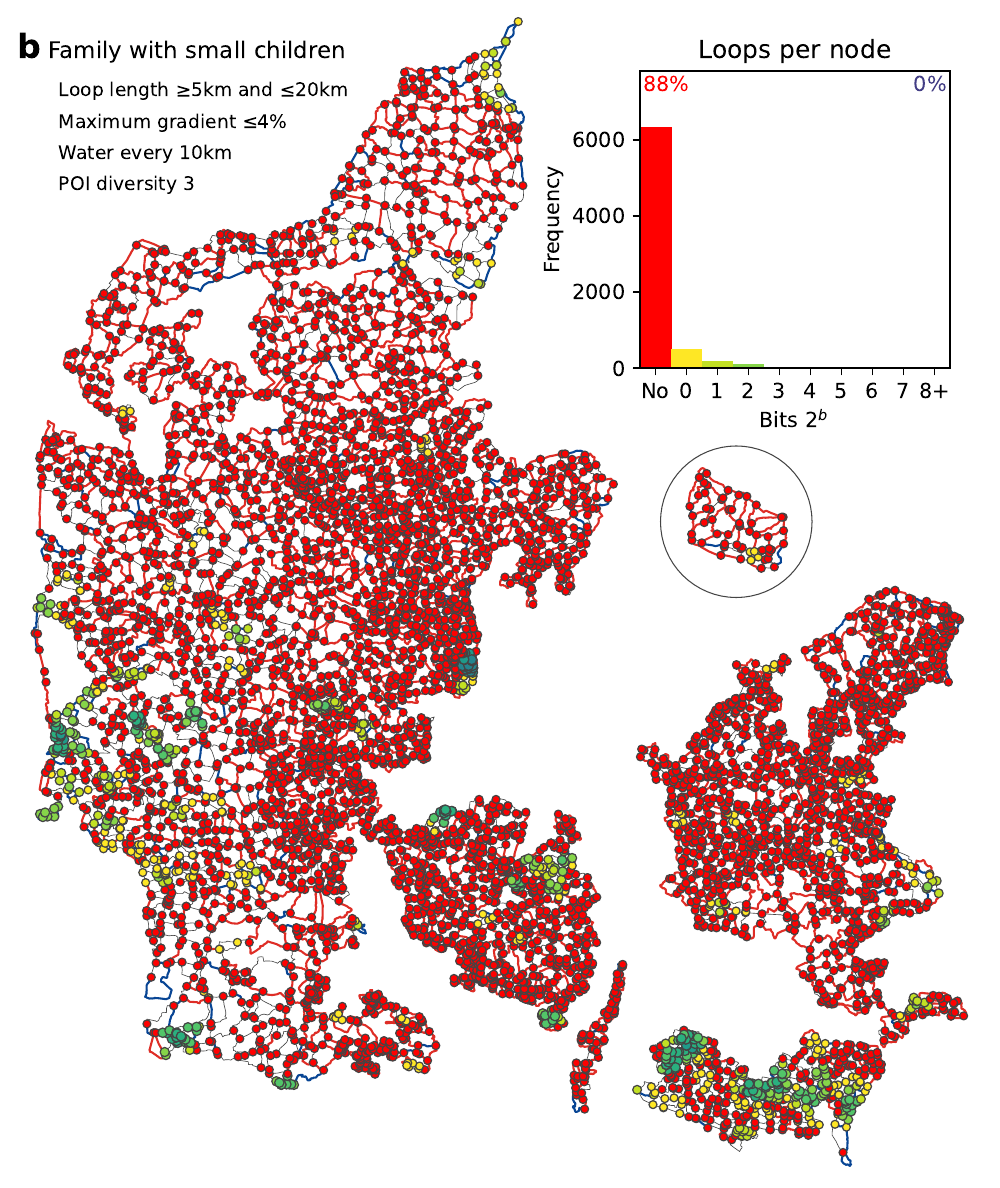}
  \end{center}
  \caption{\textbf{Loops per node for the scenario ``Family with small children''.} Adding restrictions on top of length and maximum gradient reduces the loops further. \textbf{a.} Also restricted by water available every $10\,\mathrm{km}$ (blue links), \textbf{b.} also restricted by POI diversity 3 (blue links).\label{fig:scenariofamilywaterpoi}}
\end{figure*}

\clearpage

\bibliography{references_cleaned}